\documentclass[12pt]{article}
\pdfoutput=1
\usepackage{amsmath}
\usepackage[round]{natbib}
\usepackage{graphicx,psfrag,epsf}
\usepackage{enumerate}
\usepackage{amssymb}
\usepackage{algorithm,algorithmic}
\usepackage{multirow}
\usepackage{setspace}
\usepackage{url}

\newcommand{\blind}{0}
\newcommand{\diff}{\mathrm{d}}

\newcommand{\figimage}[4]{\begin{figure}[tbp]\centering\includegraphics[width=#4 cm]{#1}\caption{#2}\label{#3}\end{figure}}
\newcommand{\figimagelf}[6]{\begin{figure}[tbp]\centering\includegraphics[width=#5 cm]{#1}\qquad\includegraphics[width=#6 cm]{#2}\caption{#3}\label{#4}\end{figure}}
 
\newcommand{\figimagethree}[5]{\begin{figure}[tbp]\begin{tabular}{cc}\begin{minipage}{0.33 \textwidth}\begin{center}\centering\includegraphics[width=5 cm]{#1}\end{center}\end{minipage}\begin{minipage}{0.33 \textwidth}\centering\includegraphics[width=5 cm]{#2}\end{minipage}\begin{minipage}{0.33 \textwidth}\centering\includegraphics[width=5 cm]{#3}\end{minipage}\end{tabular}\caption{#4}\label{#5}\end{figure}}

\begin{document}
\bibliographystyle{plainnat}
\bibpunct[:]{(}{)}{,}{a}{}{,}

\def\spacingset#1{\renewcommand{\baselinestretch}%
{#1}\small\normalsize} \spacingset{1}

\if0\blind
{
  \title{\bf Real-time Linear Operator Construction and State Estimation with the Kalman Filter}
  \author{Tsuyoshi Ishizone \thanks{
    The authors gratefully acknowledge JST, PRESTO, JPMJPR1774 and Meiji University Special Research Project.}\hspace{.2cm}\\
    Graduate School of Advanced Mathematical Sciences, Meiji University\\
    and \\
    Kazuyuki Nakamura \\
    Department of Interdisciplinary Mathematical Sciences, Meiji University\\JST, PRESTO}
  \date{}
  \maketitle
} \fi

\if1\blind
{
  \bigskip
  \bigskip
  \bigskip
  \begin{center}
    {\LARGE\bf Real-time Modeling of a Linear Gaussian State Space Model}
\end{center}
  \medskip
} \fi

\bigskip
\begin{abstract}
The Kalman filter is the most powerful tool for estimation of the states of a linear Gaussian system. In addition, using this method, an expectation maximization algorithm can be used to estimate the parameters of the model. However, this algorithm cannot function in real time. Thus, we propose a new method that can be used to estimate the transition matrices and the states of the system in real time. The proposed method uses three ideas: estimation in an observation space, a time-invariant interval, and an online learning framework. Applied to damped oscillation model, we have obtained extraordinary performance to estimate the matrices. In addition, by introducing localization and spatial uniformity to the proposed method, we have demonstrated that noise can be reduced in high-dimensional spatio-temporal data. Moreover, the proposed method has potential for use in areas such as weather forecasting and vector field analysis.
\end{abstract}

\noindent%

\spacingset{1.45}

\section{INTRODUCTION}
\label{sec:introduction}
A quick tool for noise reduction and short-term prediction is important for areas such as weather forecasting and adjusting a scanning probe microscope (SPM). In weather forecasting, engineers require a fast denoising method in order to use the results for instantaneous forecasting. Since SPMs require adjustment of some parameters for use, a method for real-time adjustment is helpful in carrying out experiments.

A number of noise reduction methods have been proposed \citep{bertalmio,elad}. In recent years, denoising methods using deep learning, such as Noise2Noise and Noise2Void, have achieved great success \citep{tian,lehtinen,krull}. However, these methods cannot predict future states.

For predicting future states, a combination of convolutional neural networks (CNNs) and recurrent neural networks (RNNs) has been proposed since ConvLSTM \citep{shi}. These methods use no model constraint and are applied to image sequence data \citep{ballas,lotter,shi2,wang}. They, however, only focus on future forecasting, and we cannot interpret the time-series model.

Focusing on the state space model, we can avoid a lack of interpretation. The Kalman filter (KF) is a powerful tool for estimate states of the linear Gaussian state space model in a parameter-given scenario \citep{kalman}. Combined with an expectation maximization (EM) algorithm, we can estimate the parameters and the states, i.e., achieve noise reduction as refined states and short-term forecasting from the constructed model \citep{shumway2}. However, the algorithm requires a great deal of calculation time and is unsuitable for real-time situation.

We therefore propose a method by which to estimate the states and the state transition of the model in real time. Using three ideas, namely, assumption of an observation transition, a time-invariant interval, and an online learning framework, the method works well for some experiments. We refer to this method as linear operator construction with the Kalman filter (LOCK).

In addition, by introducing localization and spatial uniformity to the proposed method, referred to respectively as local LOCK (LLOCK) and spatially uniform LOCK (SLOCK), we can apply the proposed methods to spatio-temporal data, such as image sequence data and grid sequence data. Applied to synthetic data, the proposed methods were found to be superior to the existing method and to be efficient in terms of calculation and memory cost.

We introduce the proposed methods in the following section. In Section \ref{sec:result}, we state the experimental results obtained using a damped oscillation model, object moving, global flow, and local stationary flow data\footnote{Our code is available at https://github.com/ZoneTsuyoshi/lock.}. Finally, we conclude the present paper in Section \ref{sec:conclusion}.

\section{METHODS}
\label{sec:method}
In this section, we first explain the notation, followed by state Kalman filter  as a benchmark method. We then propose new methods in order to estimate the states of the system and the transition matrices in real time.

\subsection{Notation}
We use bold small letters, e.g., $\boldsymbol{x}$, to denote the variable as vector. The set of natural numbers is denoted as $\mathbb{N}$, and we denote as $\mathbb{N}_n$ as the set of natural numbers that are less than or equal to $n$, i.e., $\mathbb{N}_n=\{1,2,\cdots,n\}$. For a matrix $M\in\mathbb{R}^{d\times d}$ and a vector $\boldsymbol{v}\in\mathbb{R}^d$ that depends on time $t$, we use $M_t$ and $\boldsymbol{v}_t$ for the time dependency. For a matrix $M\in\mathbb{R}^{d\times d}$ and a vector $\boldsymbol{v}\in\mathbb{R}^d$, we use $M_{ij}$ and $v_i$ to denote the $(i,j)$-th element of the matrix and the $i$-th element of the vector, respectively. Moreover, for a matrix $M\in\mathbb{R}^{d\times d}$ and a vector $\boldsymbol{v}\in\mathbb{N}^n\ (n\le d)$, $M_{\boldsymbol{v},\boldsymbol{v}}$ denotes the matrix
\[
\begin{pmatrix}
M_{v_1v_1}&\cdots&M_{v_1v_n}\\
\vdots&\ddots&\vdots\\
M_{v_nv_1}&\cdots&M_{v_nv_n}
\end{pmatrix}
\in\mathbb{R}^{n\times n}.
\]
Similarly, for vectors $\boldsymbol{v}\in\mathbb{R}^d$ and $\boldsymbol{u}\in\mathbb{N}^n\ (n\le d)$, $\boldsymbol{v}_{\boldsymbol{u}}$ denotes the vector $(\boldsymbol{v}_{u_1},\cdots,\boldsymbol{v}_{u_n})^T\in\mathbb{R}^n$. Moreover, for a set $A\subset\mathbb{R}$ that satisfies $|A|=n$, $\mathrm{vec}(A)$ denotes the vector, the $i$-th element of which corresponds to $i$-th smallest value of the matrix.

\subsection{Kalman Filter}
\label{subsec:kf}
Consider the state space model
\begin{align}
\boldsymbol{x}_t&=f_t(\boldsymbol{x}_{t-1})+\boldsymbol{v}_t,\label{eq:ssm1}\\
\boldsymbol{y}_t&=h_t(\boldsymbol{x}_t)+\boldsymbol{w}_t,\label{eq:ssm2}
\end{align}
where $\boldsymbol{x}_t\in\mathbb{R}^m$, $\boldsymbol{y}_t\in\mathbb{R}^l$, $\boldsymbol{v}_t\in\mathbb{R}^m$, and $\boldsymbol{w}_t\in\mathbb{R}^l$ represent the state of the system, the observation, the system noise, and the observation noise, respectively, at a given time $t$. In the linear scenario, using
\begin{align}
f_t(\boldsymbol{x}_{t-1})&=F_t\boldsymbol{x}_{t-1},\\
h_t(\boldsymbol{x_t})&=H_t\boldsymbol{x}_t,
\end{align}
we can obtain
\begin{align}
\boldsymbol{x}_t&=F_t\boldsymbol{x}_{t-1}+\boldsymbol{v}_t,\label{eq:lssm1}\\
\boldsymbol{y}_t&=H_t\boldsymbol{x}_t+\boldsymbol{w}_t,\label{eq:lssm2}
\end{align}
from equations \eqref{eq:ssm1} and \eqref{eq:ssm2}. Suppose the system noise and the observation noise follow a Gaussian distribution with zero mean, i.e., $\boldsymbol{v_t}\sim N(\boldsymbol{0},Q_t)$ and $\boldsymbol{w}_t\sim N(\boldsymbol{0},R_t)$. Under this assumption, we can apply the Kalman filter (KF) to the estimation of the state vectors.

The filter uses the following update formula
\begin{align}
\boldsymbol{x}_{t|t-1}&=F_t\boldsymbol{x}_{t-1|t-1},\label{eq:kfp1}\\
V_{t|t-1}&=F_tV_{t-1|t-1}F_t^T+Q_t,\label{eq:kfp2}\\
K_t&=V_{t|t-1}H_t^T(H_tV_{t|t-1}H_t^T+R_t)^{-1},\label{eq:kff1}\\
\boldsymbol{x}_{t|t}&=\boldsymbol{x}_{t|t-1}+K_t(\boldsymbol{y}_t-H_t\boldsymbol{x}_{t|t-1}),\label{eq:kff2}\\
V_{t|t}&=V_{t|t-1}-K_tH_tV_{t|t-1},\label{eq:kff3}
\end{align}
where $\boldsymbol{x}_{t|s}$ and $V_{t|s}$ represent the estimated value and the covariance, respectively, of the state at time $t$ given observation $\{\boldsymbol{y}_1,\cdots,\boldsymbol{y}_s\}$. 

The expectation maximization (EM) algorithm is one of the most powerful tools for estimation of the transition matrix. However, this algorithm is unsuitable for a real-time situation due to the requirement for a sequence of smoothed values of  $\{\boldsymbol{x}_{t|T}\}$ and $\{V_{t|T}\}$, where $T$ is final time-step of time-series data. We can consider an ad hoc strategy, namely, expectation maximization KF (EMKF), which applies the algorithm every $\tau$ timesteps. However, in the next section, we show that the proposed method only works for a limited situation. Thus, the following sections propose new methods for estimation of the transition matrices and the states in real time.

\subsection{Linear Operator Construction with Kalman Filter (LOCK)}
\label{subsec:sukf}
In order to overcome the real-time estimation problem, we propose a new method referred to as linear operator construction with the Kalman filter (LOCK) by introducing three ideas: assumption of an observation transition, a time-invariant interval, and an online learning framework.

First, we introduce the assumption of a linear and Gaussian observation transition
\begin{equation}
\boldsymbol{y}_t=G_t\boldsymbol{y}_{t-1}+\boldsymbol{u}_t,\label{eq:lossm}
\end{equation}
where $\boldsymbol{u}_t\in\mathbb{R}^l$ is the noise of the observation transition at time $t$ and follows Gaussian distribution $N(\boldsymbol{0},S_t)$. If the observation matrices are regular, this assumption is satisfied by 
\begin{align}
G_t&=H_tF_tH_{t-1}^{-1},\\
S_t&=H_{t-1}^{-T}F_t^TH_t^TR_{t-1}H_tF_tH_{t-1}^{-1}+H_t^TQ_tH_t+R_t.
\end{align}
Figure \ref{fig:sukf} shows these relationships.

Using equations \eqref{eq:lssm1}, \eqref{eq:lssm2}, and \eqref{eq:lossm}, we can obtain
\begin{equation}
\boldsymbol{y}_t=H_t(F_t\boldsymbol{x}_{t-1}+\boldsymbol{v}_t)+\boldsymbol{w}_t=G_t(H_{t-1}\boldsymbol{x}_{t-1}+\boldsymbol{w}_{t-1})+\boldsymbol{u}_t.
\end{equation}
Then, by taking the expectation, the following relation is satisfied:
\begin{equation}
(H_tF_t-G_tH_{t-1})E[\boldsymbol{x}_{t-1}]=0.
\end{equation}
Therefore, we can obtain an unbiased estimator
\begin{equation}
\hat F_t=H_t^{-}G_tH_{t-1},
\end{equation}
where $H_t^-$ and $\hat F_t$ indicate the Moore-Penrose type pseudo-inverse matrix of $H_t$ and the estimated value of $F_t$, respectively.

In addition, suppose these matrices are time-invariant in interval $\tau$, then, we can obtain 
\begin{equation}
\hat G=Y_tY_{t-1}^-,
\end{equation}
where $Y_t=(\boldsymbol{y}_{t-\tau+1},\ \cdots\, \boldsymbol{y}_t)$ is a matrix composed of the observation vectors from time $t-\tau+1$ to time $t$. Therefore, we can obtain the estimation algorithm
\begin{equation}
\hat F=H^-Y_tY_{t-1}^-H.
\end{equation}

Moreover, based on an idea incorporated from online learning \citep{bishop}, we introduce parameters in order to treat outliers in the observation
\begin{gather}
F_t=F_{t-\tau}-\eta\ \mathrm{crop}(F_{t-\tau}-\hat F,-c,c),\label{eq:sukf}\\
\left(\mathrm{crop}(F,a,b)\right)_{ij}=\min\{\max\{F_{ij},a\},b\},
\end{gather}
where $\eta$ and $c$ are the learning rate and the cutoff distance, respectively, which control the maximum amount of difference between the old estimate and the new estimate. Algorithm \ref{alg:sukf} summarizes the LOCK method.

\figimage{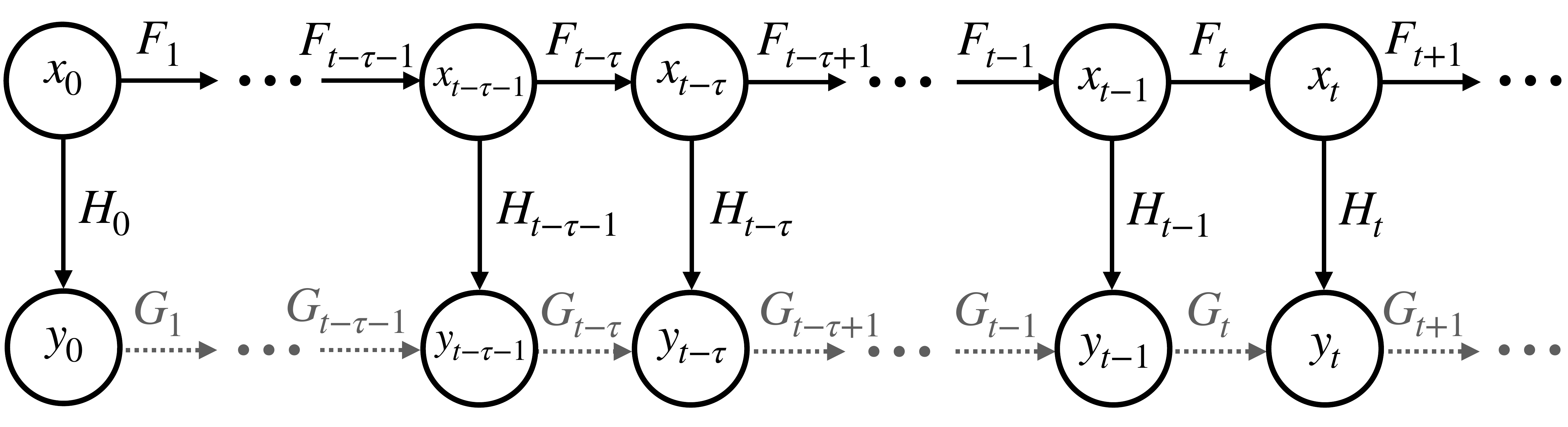}{Schematic representation of the relationships among \eqref{eq:lssm1}, \eqref{eq:lssm2}, and \eqref{eq:lossm}. The state at time $t-1$ decides the observation at time $t-1$ and the state at time $t$ by $H_{t-1}$ and $F_t$, respectively. Moreover, the observation at time $t-1$ decides the observation at time $t$ by $G_t$.}{fig:sukf}{14}

\begin{algorithm}[tbp]
\caption{Algorithm of LOCK}
\label{alg:sukf}
\begin{algorithmic}
\REQUIRE $\boldsymbol{x}_{0|0},V_{0|0},\eta,c,\tau$
\STATE Set initial time $t=0$
\WHILE{observation at a time $t$ $\boldsymbol{y}_t$ exists}
\STATE Calculate $\boldsymbol{x}_{t|t-1},V_{t|t-1}$ by equations \eqref{eq:kfp1} and \eqref{eq:kfp2}
\STATE Calculate $K_t,\boldsymbol{x}_{t|t},V_{t|t}$ by equations \eqref{eq:kff1},\eqref{eq:kff2}, and \eqref{eq:kff3}
\IF{$t\equiv0\pmod\tau$}
\STATE Update $F$ by the equation \eqref{eq:sukf}
\ENDIF
\STATE $t\leftarrow t+1$
\ENDWHILE
\end{algorithmic}
\end{algorithm}

\subsection{Local LOCK (LLOCK)}
\label{subsec:lsukf}
We introduced the LOCK method in the previous section. However, this method cannot function properly for the case in which $\tau<l$, where $\tau$ and $l$ are the update interval and the dimension of the observation, respectively. In particular, applying LOCK to movie data of large dimension, due to the requirement for a large $\tau$, is unrealistic for a real-time situation. Hence, we propose the local LOCK (LLOCK) method, which adopts localization in order to overcome this problem \citep{hunt}. 

Consider a lattice point in spatio-temporal data. The variable of the point at time $t$ is affected only by the neighborhood of the point at time $t-1$. This localization can reduce the effective observation dimension. In other words, in order to update the $(i,j)$-th element of the transition matrix, we can use only variables of the observation in the vicinity of points $i$ and $j$. 

We define a localization matrix $L=A+I$, where $A$ is an adjacency matrix
\begin{equation}
A_{ij}=\left\{\begin{array}{cl}1,&(i\text{ th variable is close to }j\text{ th variable}),\\0,&(else).\end{array}\right.
\end{equation}
Using this matrix, we can obtain
\begin{gather}
\hat G_{ij}=\left\{(Y_t)_{\boldsymbol{k},\boldsymbol{k}}\left((Y_{t-1})_{\boldsymbol{k},\boldsymbol{k}}\right)^-\right\}_{l_il_j},\\
\boldsymbol{k}=\mathrm{vec}(\{k\in\mathbb{N}_l\mid L_{ik}+L_{jk}>0\}),\\
l_i=|\{k\in\mathbb{N}_i\mid L_{i,k}+L_{j,k}>0\}|,
\end{gather}
where $l_i$ and $l_j$ are local indices of points $i$ and $j$, respectively. Algorithm \ref{alg:lsukf} summarizes the local calculation of the observation transition matrix $G$.

\begin{algorithm}[tbp]
\caption{Local calculation of the observation transition matrix}
\label{alg:lsukf}
\begin{algorithmic}
\REQUIRE $L,\{\boldsymbol{y_s}\}_{s=t-\tau}^t$
\FOR{$i\in\{1,\cdots,l\}$}
\FOR{$j\in\{1,\cdots,l\}$}
\IF{$L_{i,j}=1$}
\STATE Construct the vector $\boldsymbol{k}$
\STATE Calculate the local indices $l_i$ and $l_j$ corresponding to the global indices $i$ and $j$
\STATE Localize $\{\boldsymbol{y}_s\}_{s=t-\tau}^t$ to $\{\boldsymbol{y}_s^{local}\}_{s=t-\tau}^t$, where $\boldsymbol{y}_s^{local}=(\boldsymbol{y}_s)_{\boldsymbol{k}}$
\STATE $Y_{t-1}^{local}\leftarrow(\boldsymbol{y}_{t-\tau}^{local},\cdots,\boldsymbol{y}_{t-1}^{local})$
\STATE $Y_{t}^{local}\leftarrow(\boldsymbol{y}_{t-\tau+1}^{local},\cdots,\boldsymbol{y}_{t}^{local})$
\STATE $\hat G^{local}\leftarrow Y_t^{local}(Y_{t-1}^{local})^-$
\STATE $\hat G_{ij}\leftarrow \hat G^{local}_{l_il_j}$
\ENDIF
\ENDFOR
\ENDFOR
\end{algorithmic}
\end{algorithm}

\subsection{Spatially Uniform LOCK (SLOCK)}
\label{subsec:psukf}
The LOCK and LLOCK methods consider the elements of the transition matrix as independent parameters, as a result, these methods treat many parameters and require a long update interval in order to avoid $\tau<l$. Thus, both methods are unsuitable for rapid change of the transition matrix. Therefore, we introduce the spatially uniform LOCK (SLOCK) method to solve this problem. This method considers multiple elements of the transition matrix as a common parameter. In other words, we assume that parameters are spatially uniform in given vector field. Since the number of parameters is smaller than that of LOCK or LLOCK, the algorithm of this method can update with a shorter interval.

First, we design a matrix $P\in(\{0\}\cup\mathbb{N})^l\times(\{0\}\cup\mathbb{N})^l$, each element of which represents a parameter number, in which if two elements are same, then the corresponding two elements of $G$ are same, i.e.,
\begin{equation}
P_{i_1j_1}=P_{i_2j_2}\Rightarrow G_{i_1j_1}=G_{i_2j_2}.
\end{equation}
In addition, similar to the localization matrix, the zero elements of $P$ correspond to the zero elements of $G$, i.e.,
\begin{equation}
P_{ij}=0\Rightarrow G_{ij}=0.
\end{equation}
If we use the numbers in ascending order, then we have $\alpha\le l^2$ unique values, excluding zero, which are $\{1,\cdots,\alpha\}$. Next, we use the vector $\boldsymbol{\theta}\in\mathbb{R}^\alpha$ to denote the values of $G$
\begin{equation}
G_{ij}=
\left\{\begin{array}{ll}
\boldsymbol{\theta}_{P_{ij}},&(P_{ij}>0),\\
0,&(P_{ij}=0).
\end{array}\right.
\end{equation}
Thanks to this notation, we can obtain 
\begin{gather}
\boldsymbol{y}_t=\Xi_{t-1}\boldsymbol{\theta},\\
(\Xi_t)_{ij}=\sum_{k\in B_{ij}}(\boldsymbol{y}_t)_k,\ \Xi\in\mathbb{R}^l\times\mathbb{R}^\alpha,\\
B_{i,j}=\{k\in\mathbb{N}_l\mid P_{ik}=j\}\ (i\in\mathbb{N}_l,j\in\mathbb{N}_\alpha),
\end{gather}
where $\boldsymbol{\theta}$ is the parameter vector. Using this equation, we can obtain the following update equations:
\begin{align}
\hat{\boldsymbol{\theta}}&=\Xi_{t-1}^-\boldsymbol{y}_t,\label{eq:psukf1}\\
\left(\hat G\right)_{ij}&=\left(\hat{\boldsymbol{\theta}}\right)_{P_{ij}},\label{eq:psukf2}\\
\hat F&=H_t^-\hat GH_{t-1}.\label{eq:psukf3}
\end{align}
Algorithm \ref{alg:psukf} summarizes this method.

\begin{algorithm}[tbp]
\caption{Estimation algorithm of the transition matrix used by SLOCK}
\label{alg:psukf}
\begin{algorithmic}
\REQUIRE $P,\{\boldsymbol{y_s}\}_{s=t-1}^t$
\FOR{$i\in\{1,\cdots,l\}$}
\FOR{$j\in\{1,\cdots,l\}$}
\STATE Construct the set $B_{ij}$
\STATE Calculate the $(i,j)$-th element of the matrix $\Xi_{t-1}$
\ENDFOR
\ENDFOR
\STATE Calculate $\hat{\boldsymbol{\theta}},\hat G,\hat F$ by equations \eqref{eq:psukf1}-\eqref{eq:psukf3}
\end{algorithmic}
\end{algorithm}

\section{RESULTS}
\label{sec:result}
We applied EMKF and LOCK to the damped oscillation model, SLOCK to object moving data and global flow data, and LLOCK to global flow data and local stationary flow data.

\subsection{Damped Oscillation Model}
\label{ssec:dom}

\subsubsection{Model Equation}
Damped oscillation model
\begin{align}
\frac{\diff x}{\diff t}&=v,\\
m\frac{\diff v}{\diff t}&=-kx-rv,
\end{align}
represents the behavior of a damped oscillator, which is subject to a damping force $rv$, where $x$ is the position of the object, $v$ is the velocity, $m$ is the mass, $k$ is the oscillation constant, and $r$ is the damping constant. Using the Euler forward method, we can obtain
\begin{align}
x_t&=x_{t-1}+\Delta t\ v_{t-1},\\
v_t&=-\frac{k}{m}\Delta t\ x_{t-1}+\left(1-\frac{r}{m}\Delta t\right)v_{t-1},
\end{align}
where $\Delta t$ is the step size. By combining $x_t$ and $v_t$ into the state vector $\boldsymbol{x}_t$, we can obtain
\begin{gather}
\boldsymbol{x}_t=F^{true}\boldsymbol{x}_{t-1},\label{eq:domm1}\\
\boldsymbol{x}_t=\begin{pmatrix}
x_t\\v_t
\end{pmatrix},\ 
F^{true}=\begin{pmatrix}
1&\Delta t\\
-k\Delta t/m&1-r\Delta t/m
\end{pmatrix},\label{eq:domm2}
\end{gather}
where $F^{true}$ is the true transition matrix of this model. Appling this model to the linear state space model, we can use the EMKF and LOCK methods.

\subsubsection{Experimental Conditions}
We conducted five experiments. In the first and second experiments, the parameters of the model are time-invariant. In the other experiments, the parameters of the model are time-variant. The five experiments use the same setting, excluding the model parameters, as shown in Table \ref{tab:dom_param}. 

In experiment 1, we first simulated the true model from $\boldsymbol{x}_0=(5.0,\ 0.0)^T$ for $T=100$ adding system noise $\boldsymbol{v}_t\sim N(\mathbf{0},0.01^2\times I)$ (referred to herein as ``true" states). Then, adding observation noise $\boldsymbol{w}_t\sim N(\mathbf{0},0.2^2\times I)$, we can obtain ``pseudo-observation" data. Then, we applied EMKF and LOCK to the observation and obtained the experimental results. Here, the initial transition matrix $F_0$ is the true one, and the initial state and the transition covariance are different from the condition for generating the true data.

Experiment 2 differs from the previous experiment with respect to the initial transition matrix and the number of simulations. The initial transition matrix is not the true one but rather follows isotropic Gaussian distribution $IN(F^{true},1.0^2\times I)$, and each element $(i,j)$ of the matrix independently follows Gaussian distribution $N(F_{ij},1.0^2)$. Regarding the number of simulations, considering randomness, we executed 100 simulations and obtained the experimental results.

The main difference between experiment 3 and experiments 1 and 2 is whether parameters are time-invariant or time-variant. Experiments 3 and 4 use the same settings as experiments 1 and 2, respectively, excluding $k$, $r$, and $\eta$ (see Table \ref{tab:dom_param}). Experiments 4 and 5 have different distributions of the initial transition matrices, which are $IN(F_0^{true},1.0^2\times I)$ and $IN(F_0^{true},0.01^2\times I)$ for experiments 4 and 5, respectively.

\begin{table}[tbp]
\caption{Parameters of the experiments using the damped oscillation model}
\label{tab:dom_param}
\centering
\begin{tabular}{l|l}\hline
damped oscillation model&$\Delta t=1.0$, $m=1.0$, final time-step $T=100$\\
state space model&$\boldsymbol{x}_0=\begin{pmatrix}6\\0\end{pmatrix}$, $H=V_0=I$, $R=0.2^2\times I$, $Q=0.01^2\begin{pmatrix}0&0\\0&(\Delta t/m)^2\end{pmatrix}$\\
LOCK&$\tau=4$, $c=0.5$\\
EMKF&$\tau=4$, $c=0.5$, \#(iteration)$=5$\\\hline
experiment 1&$k=0.5$, $r=0.52$, $\eta=0.6$, $F_0=F^{true}$\\
experiment 2&$k=0.5$, $r=0.52$, $\eta=0.6$, $F_0\sim IN(F^{true},1.0^2\times I)$\\
\multirow{2}{*}{experiment 3}&$k(t)=0.65t/T+0.35(1-t/T)$, $r(t)=0.37t/T+0.67(1-t/T)$,\\
&$\eta=0.8$, $F_0=F^{true}_0$\\
\multirow{2}{*}{experiment 4}&$k(t)=0.65t/T+0.35(1-t/T)$, $r(t)=0.37t/T+0.67(1-t/T)$,\\
&$\eta=0.8$, $F_0\sim IN(F^{true}_0,1.0^2\times I)$\\
\multirow{2}{*}{experiment 5}&$k(t)=0.65t/T+0.35(1-t/T)$, $r(t)=0.37t/T+0.67(1-t/T)$,\\
&$\eta=0.8$, $F_0\sim IN(F^{true}_0,0.01^2\times I)$\\\hline
\end{tabular}
\end{table}

\subsubsection{Experimental Results and Considerations}
Figure \ref{fig:dom_state} illustrates the time transition of the true states, the observation, and the estimated results used by LOCK and EMKF in experiment 1. From this figure, LOCK and EMKF function well in terms of the estimation of the states.

Figure \ref{fig:cdom_state} shows similar results to those for experiment 3. Although the estimated results for LOCK sometimes extend away from the true states, both methods work well overall for state estimation of the time-variant model.

In addition, Figures \ref{fig:cdom_lock_mat} and \ref{fig:cdom_emkf_mat} show the time transition of the estimated matrices used by LOCK and EMKF, respectively. The four panels correspond to the four elements of the matrices; e.g., the upper-left panel corresponds to $F_{11}$. From Figure \ref{fig:cdom_lock_mat}, the estimated matrices of LOCK are similar to the true matrix, even if the initial transition matrices are far from the true matrix. On the other hand, as shown in Figure \ref{fig:cdom_emkf_mat}, EMKF works only when the initial transition matrix is similar to the true matrix. In the real world, since the initial transition matrix is often unknown, LOCK is expected to be better than EMKF, from the perspective of approaching the true matrix.

Similar results for experiments 2 and 5 are presented in Appendix \ref{sec:dom}.

\figimage{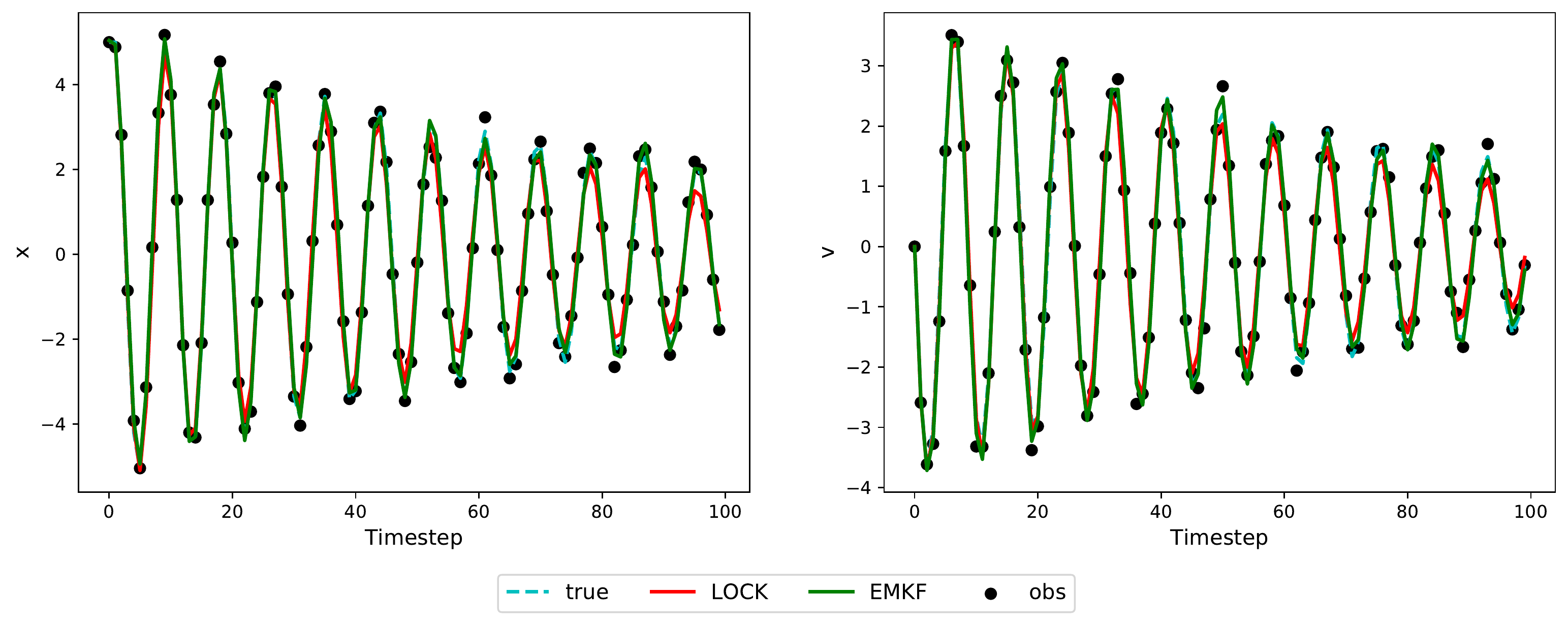}{Time transition of the true states, the observations, and the estimated results used by LOCK and EMKF in experiment 1.}{fig:dom_state}{11}
\figimage{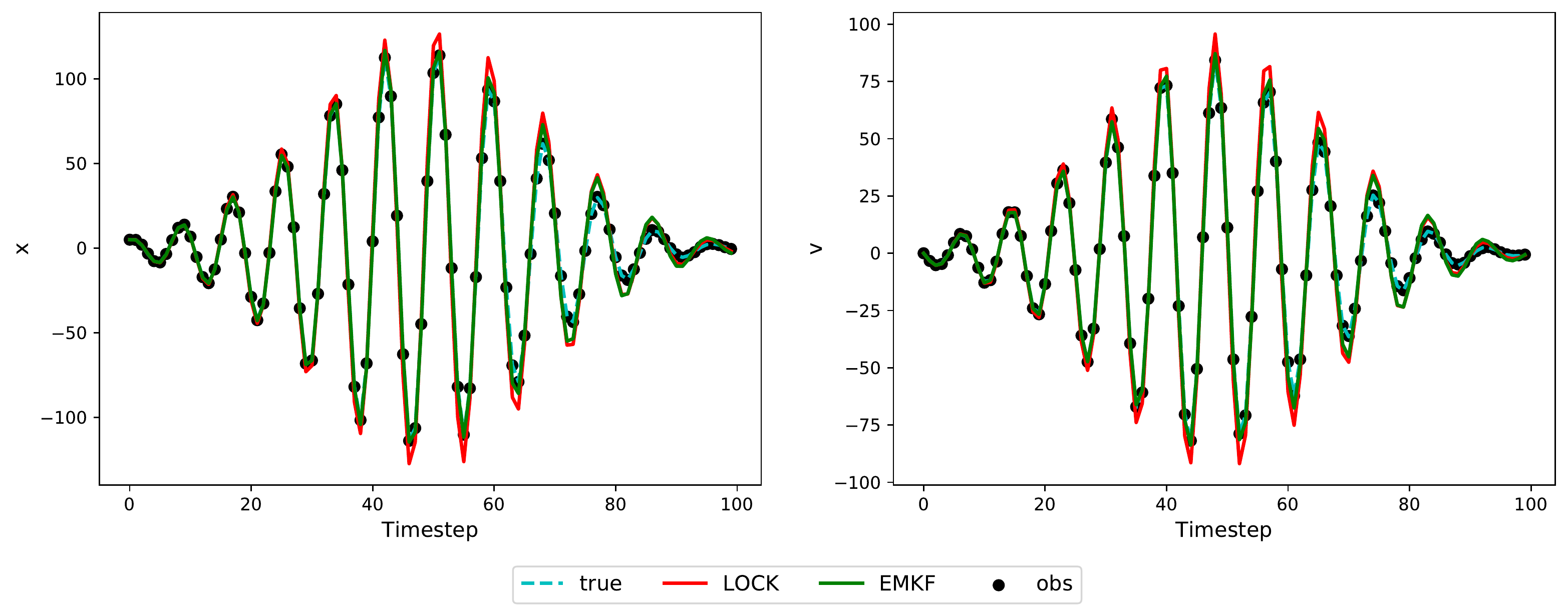}{Time transition of the true states, the observations, and the estimated results used by LOCK and EMKF in experiment 3.}{fig:cdom_state}{11}
\figimage{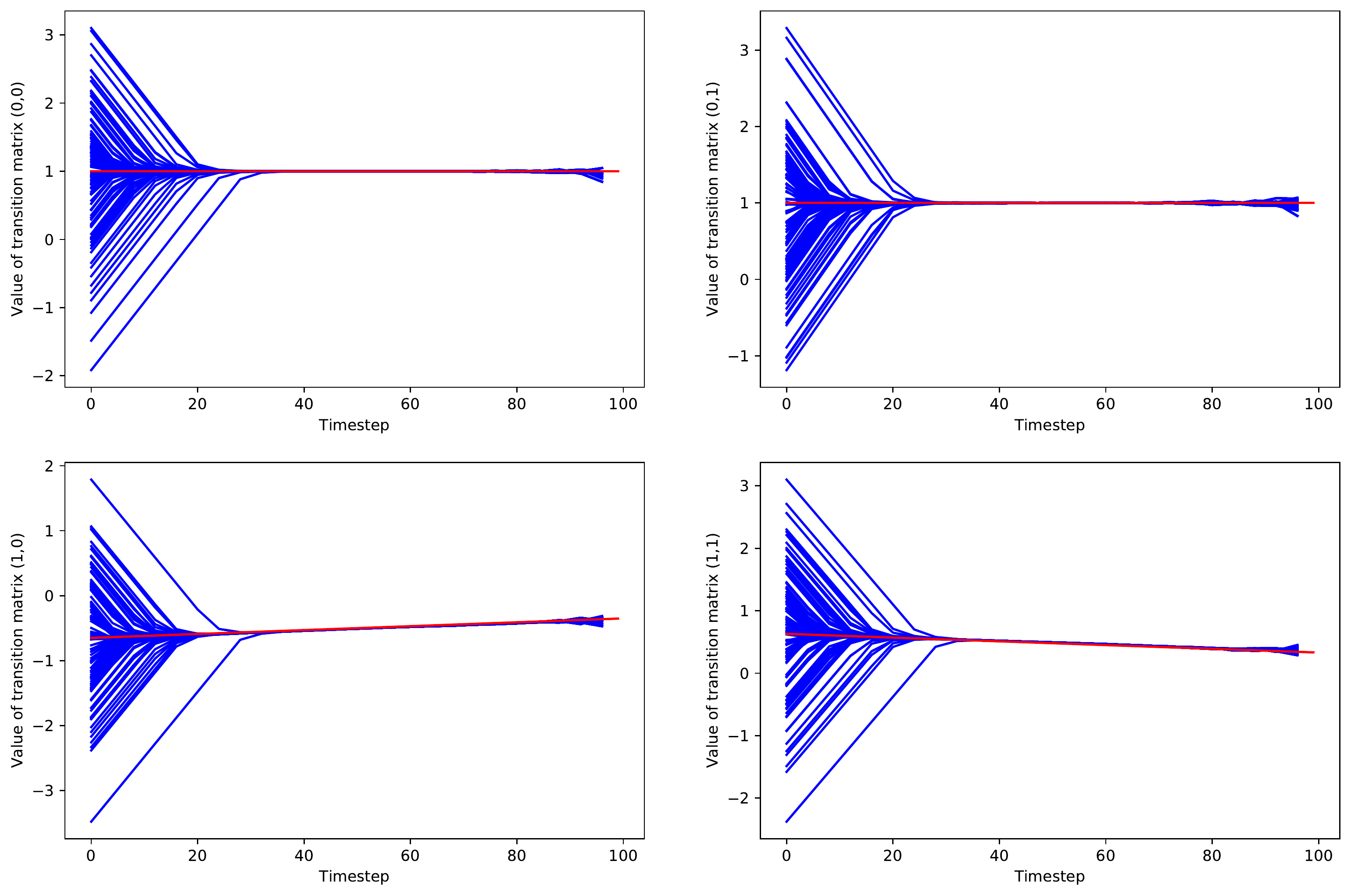}{Time transition of the true transition matrices ({\it red}) and the estimated transition matrices by LOCK ({\it blue}) in experiment 4. The four panels correspond to the four elements of the matrices; e.g., the upper-left panel corresponds to $F_{11}$.}{fig:cdom_lock_mat}{11}
\figimage{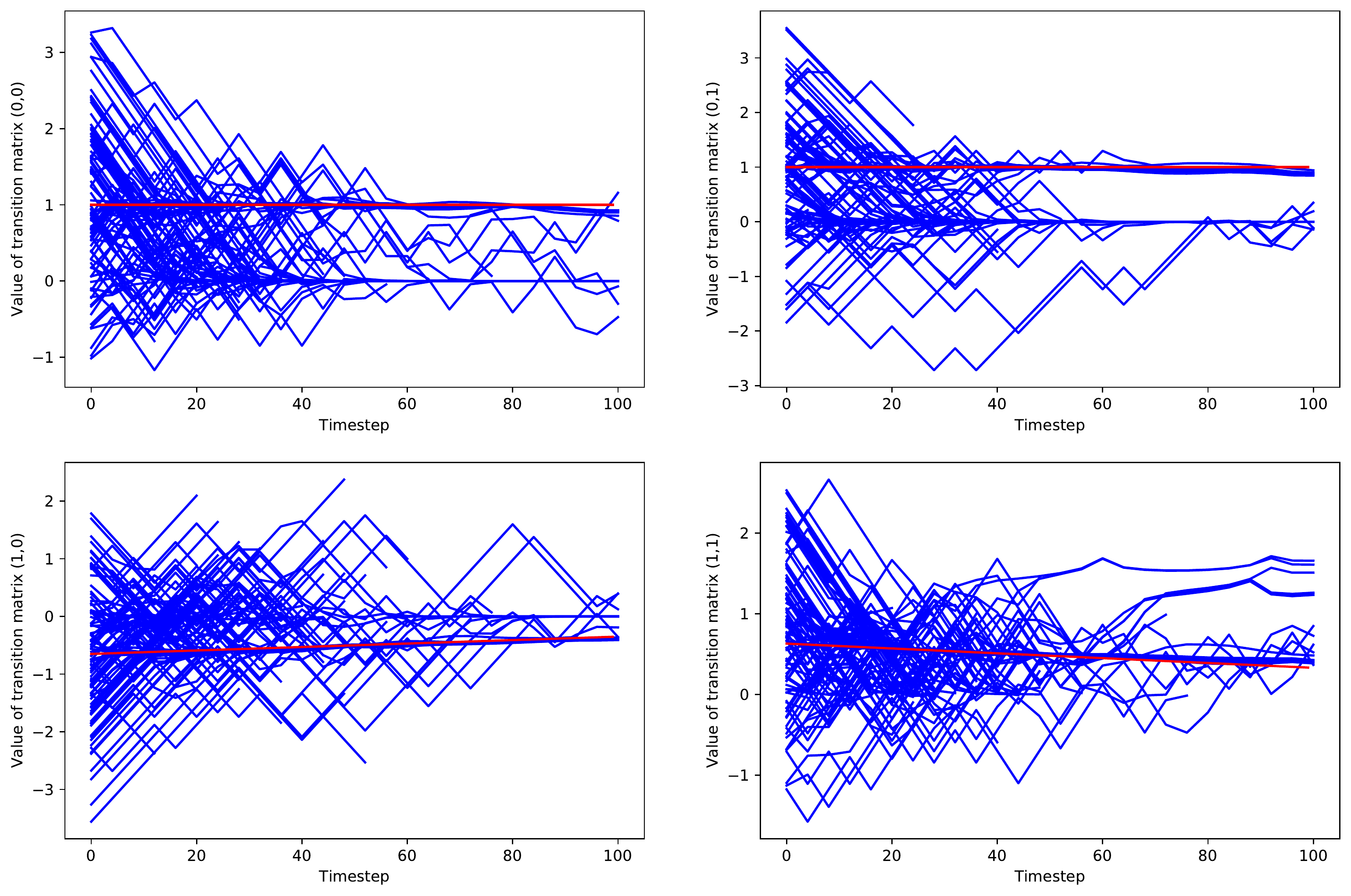}{Time transition of the true transition matrices ({\it red}) and the estimated transition matrices by EMKF ({\it blue}) in experiment 4. The four panels correspond to the four elements of the matrices; e.g., the upper-left panel corresponds to $F_{11}$.}{fig:cdom_emkf_mat}{11}

\subsection{Object Moving}
\label{ssec:mt}
\subsubsection{Experimental condition}
We generated movie data in which an object moves various directions for $T=100$, referred to herein as ``object moving" data. Details are presented in Appendix \ref{ssec:mt_data}. After generating the true data, we added Gaussian noise $N(0,20^2)$. Figure \ref{fig:mt_ex} shows the pseudo-observation at $t=0,5,10,15,20$. The directions of transition often change and summarized in Figure \ref{fig:mt_data}. In addition, we used the ``adjacency distance" as the neighbors considered during the localization phase, as shown in Figure \ref{fig:ad_dist}. We set the parameters of SLOCK as $\tau=1$, $\eta=c=1.0$, and $d=1$ and the state space model given by $F_0=H=V_0=I$ and $Q=R=0.2^2\times I$.

\figimage{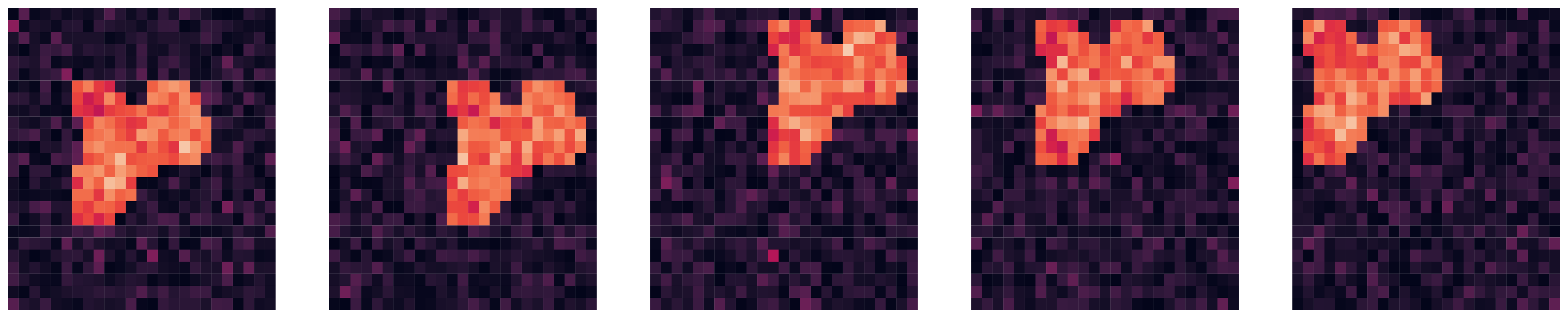}{Observation of object moving data at time $t=0,5,10,15,20$.}{fig:mt_ex}{15}
\figimage{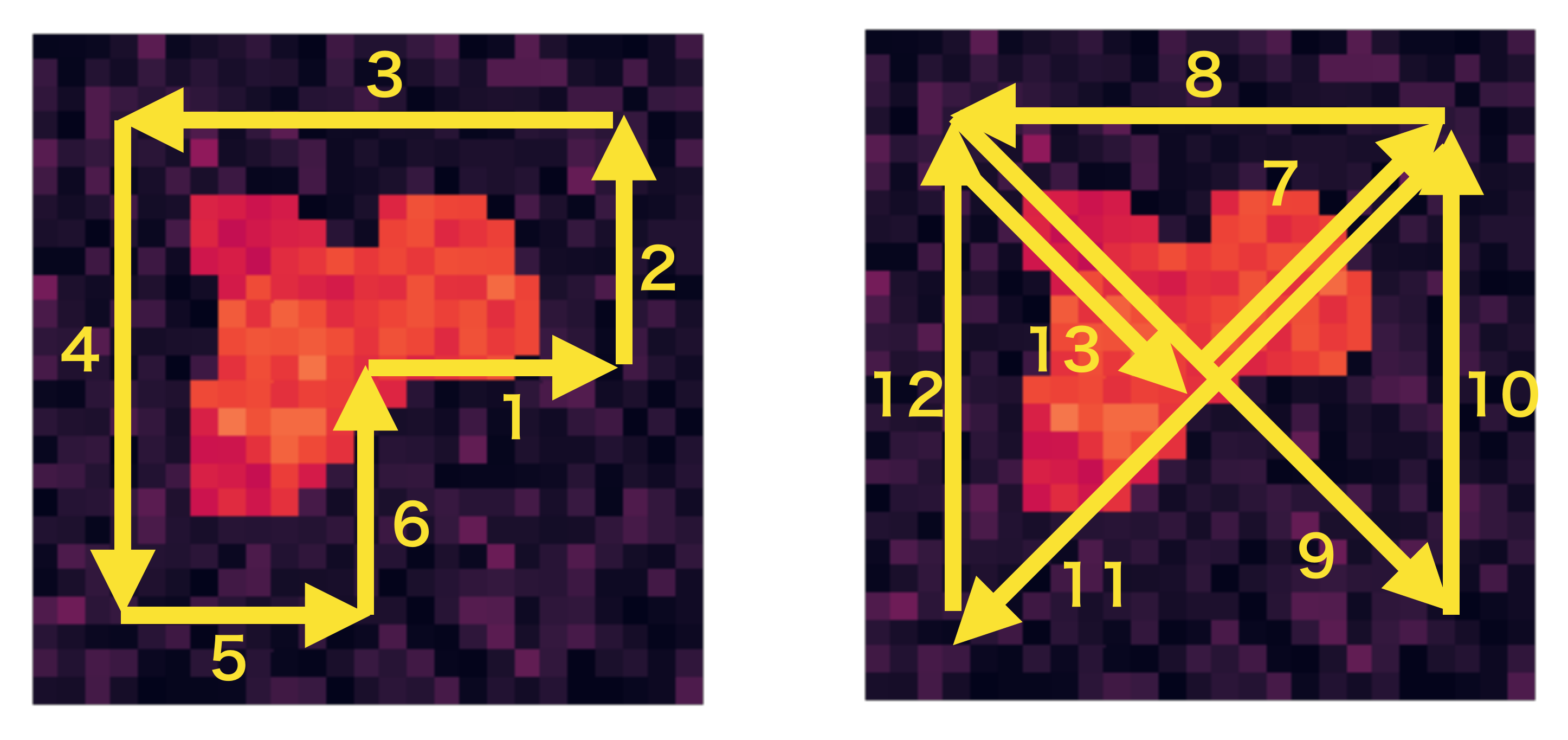}{Directions of transition of the object at each interval.}{fig:mt_data}{12}
\figimage{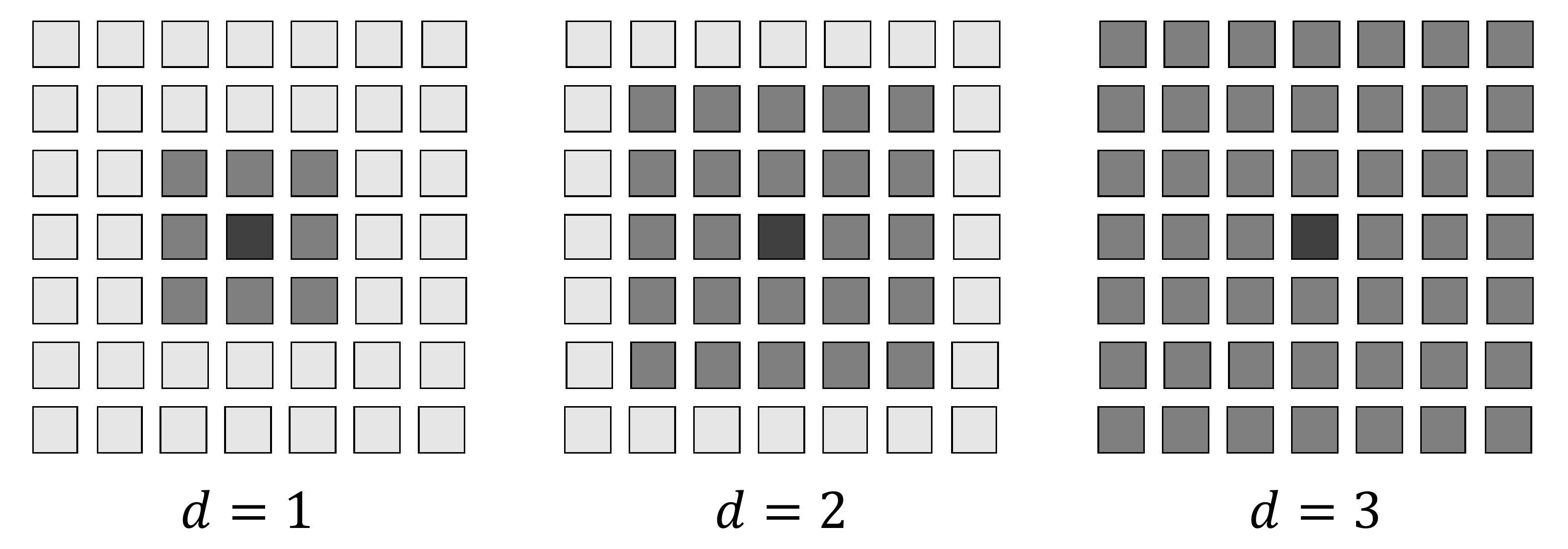}{Relationship between the adjacency distance $d$ and the localized range. For example, $d=1$ represents that a value at the $i$-th grid at time $t$ is only affected by neighbor $3\times3$ values at time $t-1$.}{fig:ad_dist}{12}

\subsubsection{Experimental Results and Considerations}
Figure \ref{fig:mt_rmse} shows the time transition of the root mean squared error (RMSE) of the observations and estimated results used by KF and SLOCK. From this figure, since the RMSE of SLOCK is lower than that of KF and the observations, this method is a powerful tool for estimating the spatially uniform model.

In addition, we calculated the RMSE for the transition matrices and executed sensitivity analysis for $d$, $\eta$, and $c$ (as shown in Appendix \ref{ssec:mt_res}).

\figimage{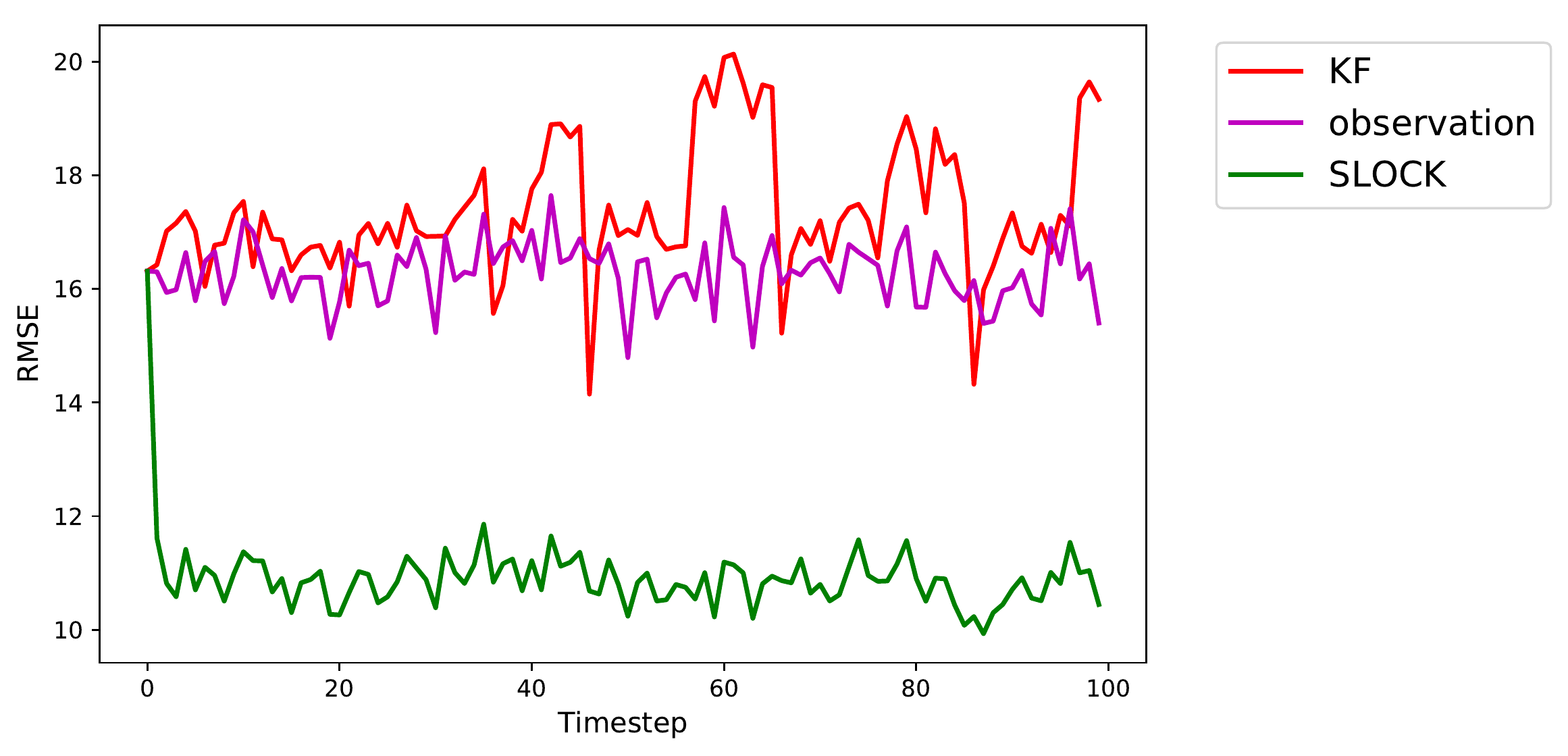}{Time transition of the RMSE of the observations and estimated results used by KF and SLOCK. }{fig:mt_rmse}{15}

\subsection{Global Flow}
\label{ssec:gf}
\subsubsection{Experimental Conditions}
We generated ``global flow" data, where various objects exist in images and move in each direction in each interval, as shown in Figure \ref{fig:gf_data}. A more detailed description of the generating process is stated in Appendix \ref{ssec:gf_data}. We added Gaussian noise $N(0,20^2)$ at each grid and obtained pseudo-observation data. We set the parameters of LLOCK to $\tau=50$, $\eta=0.8$, $c=1.0$, and $d=1$ and the state space model given by $F_0=H=V_0=I$ and $Q=R=0.2^2\times I$.

\figimage{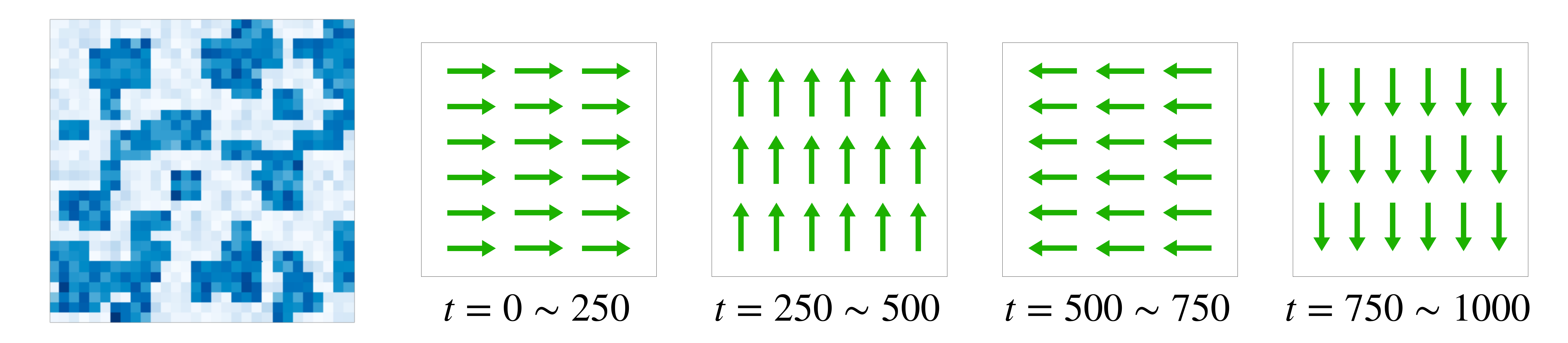}{(Left) Initial image of the global flow data. (Other panels) Flows of objects at each interval.}{fig:gf_data}{15}

\subsubsection{Experimental Results and Considerations}
Figure \ref{fig:gf_rmse} shows the time transition of the RMSE of the observations and estimated results used by KF and LLOCK. As shown in this figure, LLOCK provides better estimated results, excluding near the changing points. In this data setting, since the changes are rapid, the RMSE of the proposed method is worse than the observation in the vicinity of the changes. 

In addition, we conducted short-term prediction by KF and LLOCK. The results are shown in Figure \ref{fig:gf_stp}. We simulate each method until $t=200$ and perform prediction thereafter, where the predicted RMSE of the observations represent the RMSE between the true state at a time and the observation at $t=200$. From this, the predictive ability of LLOCK is shown to be superior to that of KF and ad hoc observation for this data.

Figures regarding the RMSE for the transition matrices and sensitivity analysis for $\tau$, $\eta$, and $c$ are shown in Appendix \ref{ssec:gf_res}.

Moreover, we measured the calculation time of updating the transition matrices for the case in which images are $10\times10$, $20\times20$, and $30\times30$. We applied SLOCK and LLOCK to data in this setting, as shown in Figure \ref{fig:ccost}. From this figure, the calculation times of both methods are low enough to be executed in a real-time situation.

In addition, we calculated ideal memory cost and ad hoc memory cost for the proposed methods and the EM algorithm. Figure \ref{fig:mcost} shows these results when we assume a float-type array. Here, ``ad hoc" means that we execute no memory-saving code, i.e., that we memorize the $l\times l$ matrices as is. However, thanks to localization and the spatially uniform assumption, we need only manage $l\times N_l$ arrays, referred to as the ``ideal" cost, where $N_l$ represents the local dimension. According to this figure, the ideal memory cost is much less than that of the EM algorithm and the methods are easier to apply.

\figimage{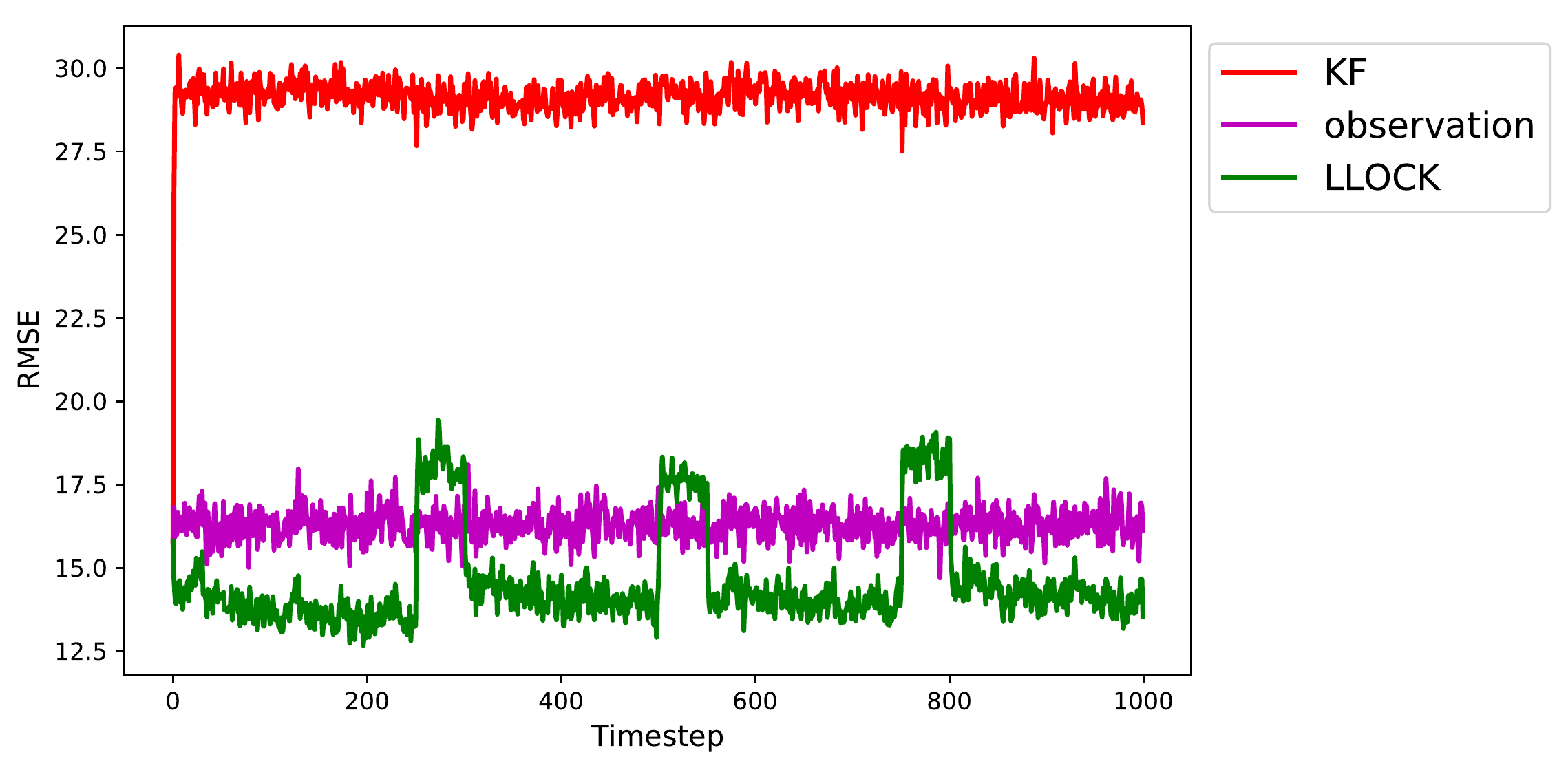}{Time transition of the RMSE of the observations and estimated results used by KF and LLOCK.}{fig:gf_rmse}{15}
\figimage{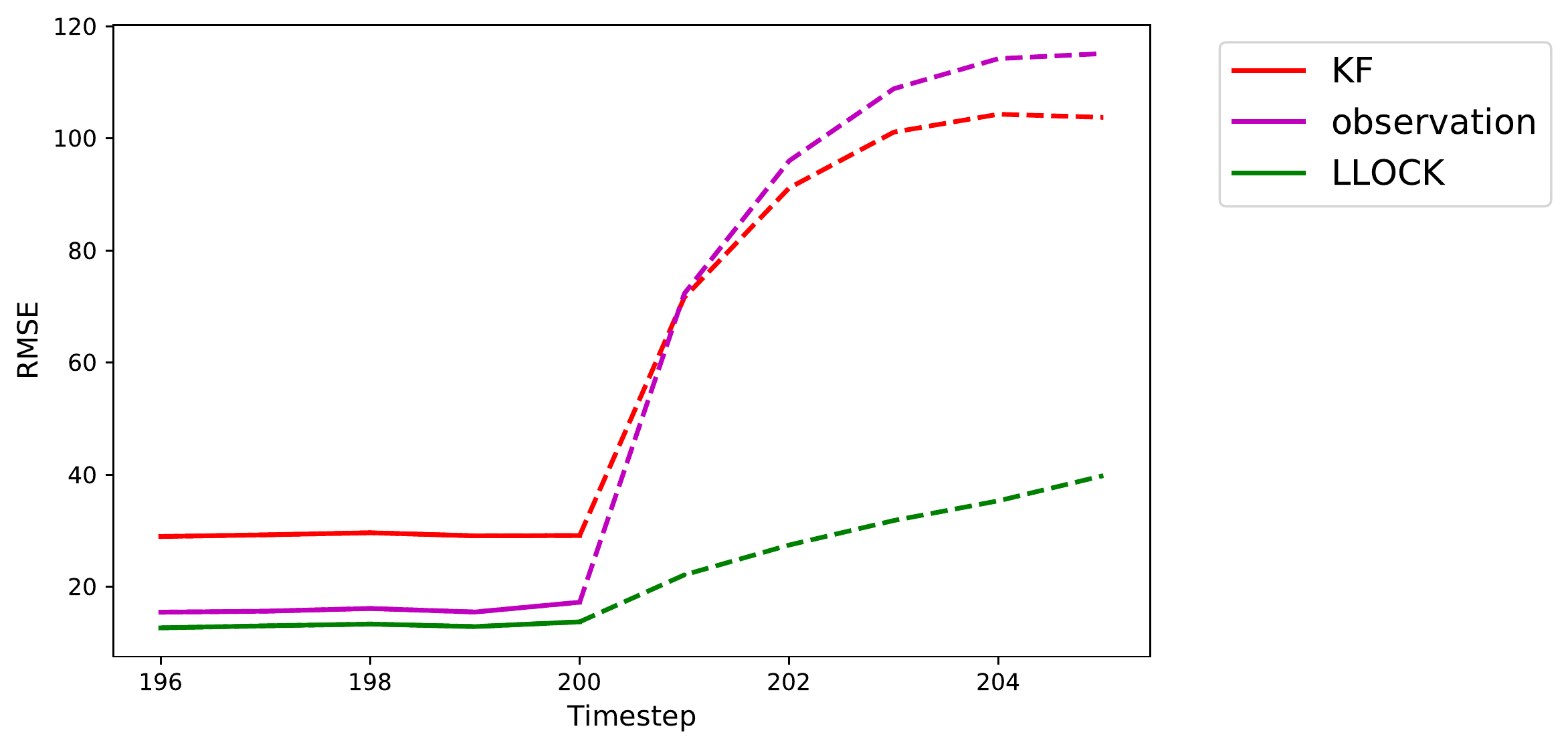}{Time transition of the RMSE of the observations and predicted results used by KF and LLOCK when we obtain data until $t=200$ and perform prediction thereafter.}{fig:gf_stp}{15}
\figimagelf{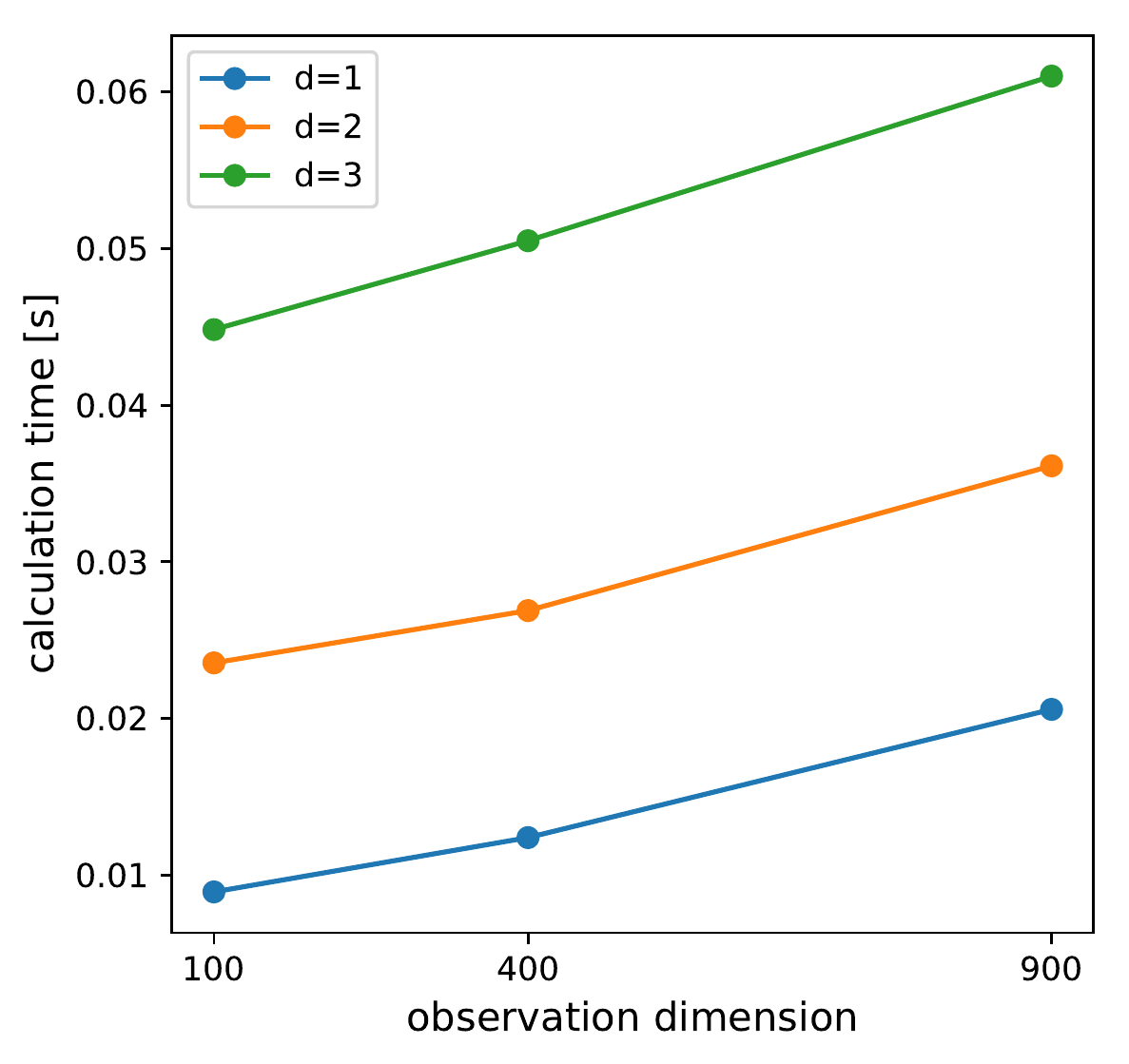}{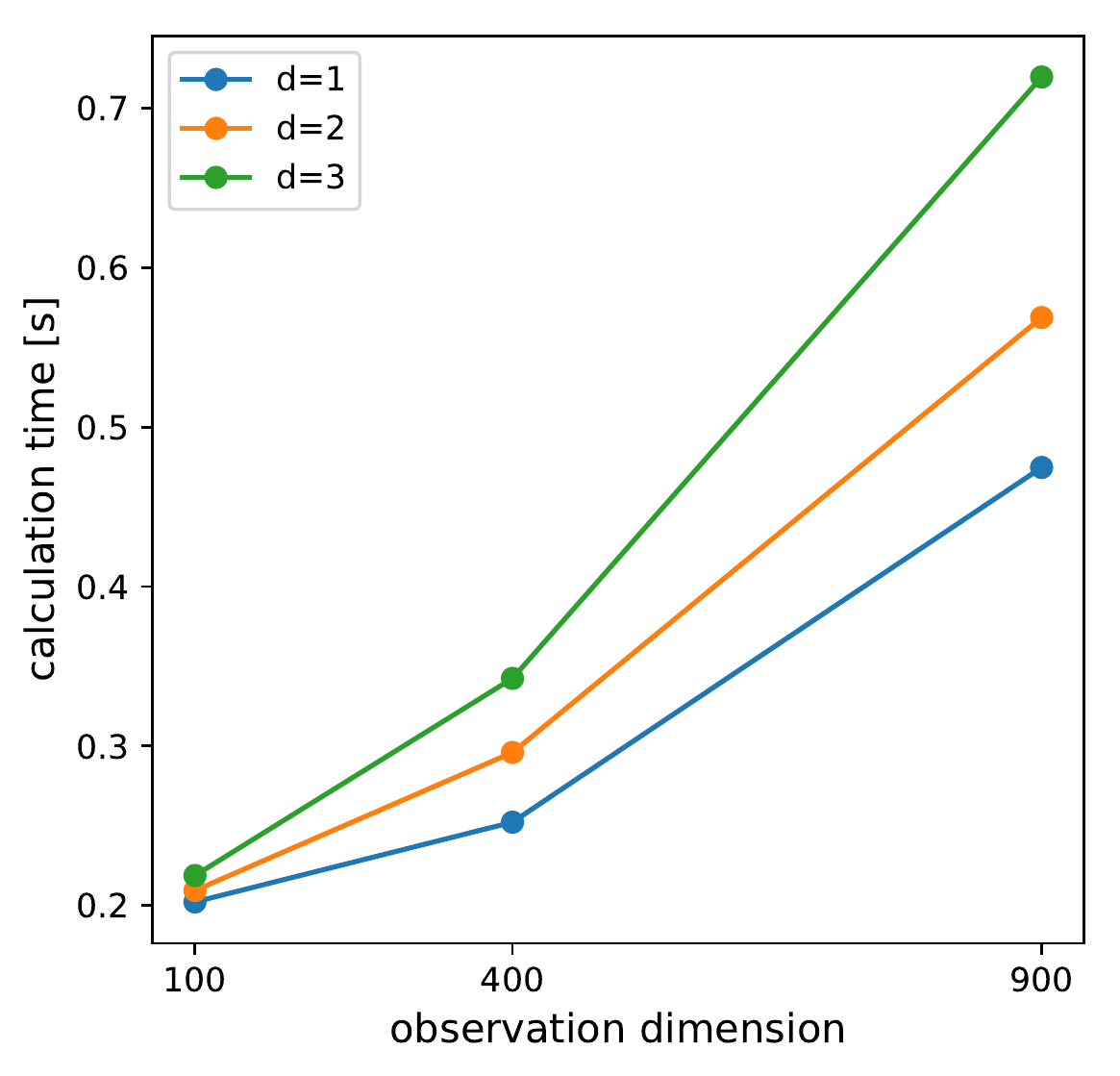}{Calculation time when the observation dimensions are varied for (Left) SLOCK and (Right) LLOCK.}{fig:ccost}{7}{7}
\figimage{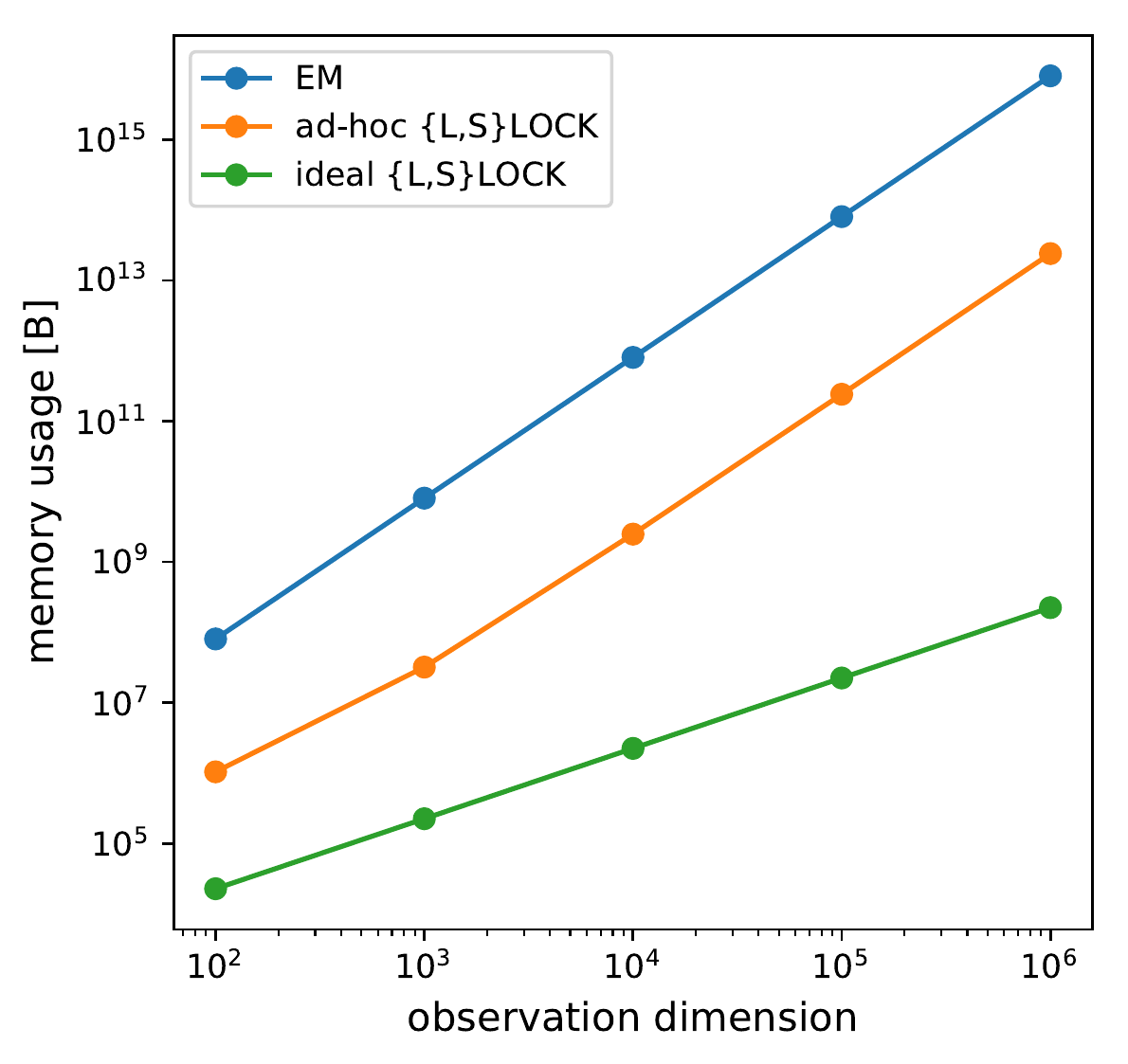}{Memory usage when observation dimensions are varied assuming a float-type array.}{fig:mcost}{7}

\subsection{Local Stationary Flow}
\label{ssec:lsf}

\subsubsection{Experimental Condition}
We generated ``local stationary flow" data for which objects spring up in the boundary and move in four directions corresponding to each field, as shown in Figure \ref{fig:lsf_data}. The left-hand panel represents the flow of the data. For example, the flow direction of the upper-left part of the image is upward. The other panels show the pseudo-observation data at $t=0$, 5, and 10. A more detailed explanation of the generating process is shown in Appendix \ref{ssec:lsf_data}. Generating the true data, we added Gaussian noise $N(0,20^2)$ and gained the observations.

We then applied LLOCK to these data to confirm that the method can capture local information. We set the parameters of LLOCK as $\tau=50$, $\eta=0.6$, $c=1.0$, and $d=1$ and the state space model given by $F_0=H=V_0=I$ and $Q=R=0.2^2\times I$.

\figimagelf{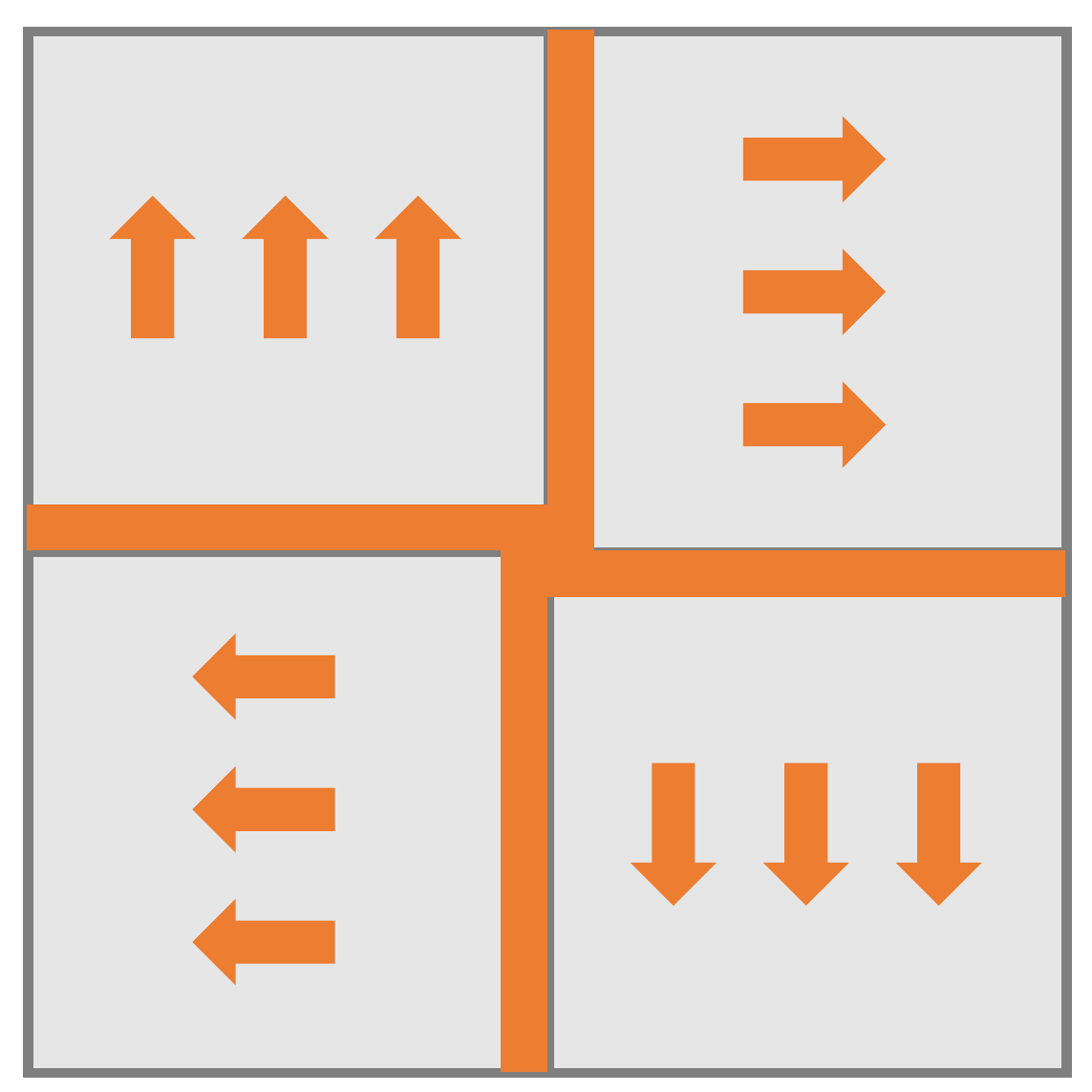}{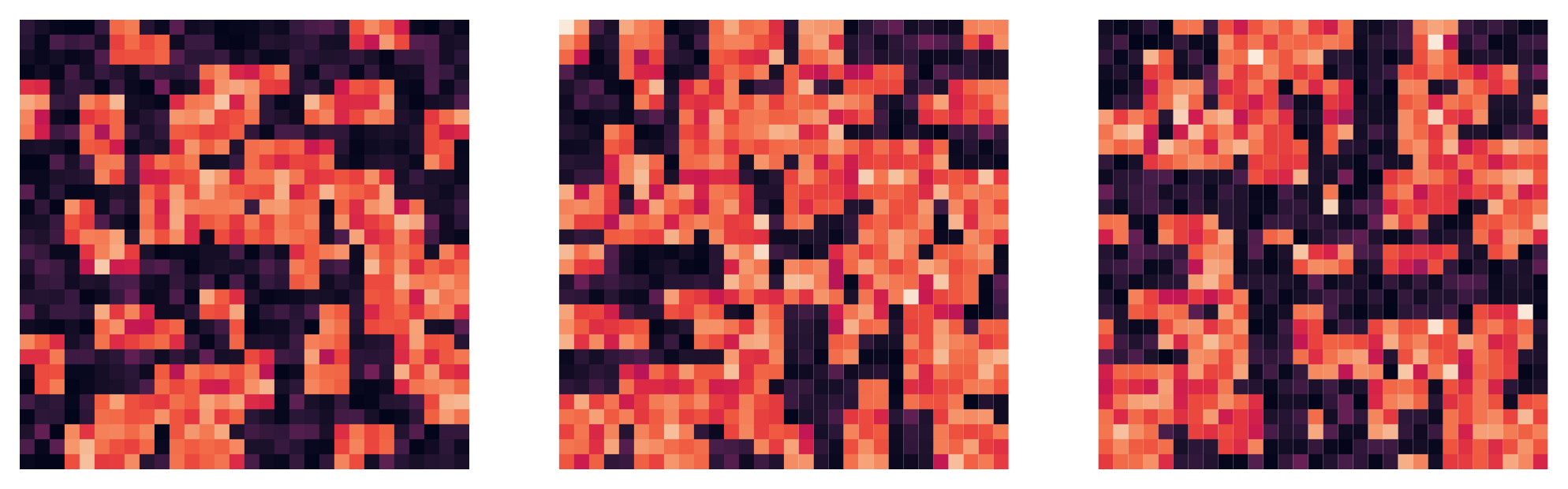}{Generation of local stationary flow data. (Left) Flow of the data. For example, the flow direction of the upper-left part of the image is upward. (Other) Pseudo-observation data at $t=0$, $5$, and $10$.}{fig:lsf_data}{4}{12}

\subsubsection{Experimental Results and Considerations}
Figure \ref{fig:lsf_rmse} shows the time transition of the RMSE of the observations and the estimated results used by KF and LLOCK. From this figure, LLOCK has a lower RMSE than that of KF and the observations. 

In addition, we performed prediction for the short period used by KF and LLOCK as shown in Figure \ref{fig:lsf_stp}. Similar to the results for global flow data, the proposed method has better performance than the other methods for short-term prediction.

The RMSE regarding the transition matrix and sensitivity analysis for $\tau$, $\eta$, and $c$ are presented in Appendix \ref{ssec:lsf_res}.

\figimage{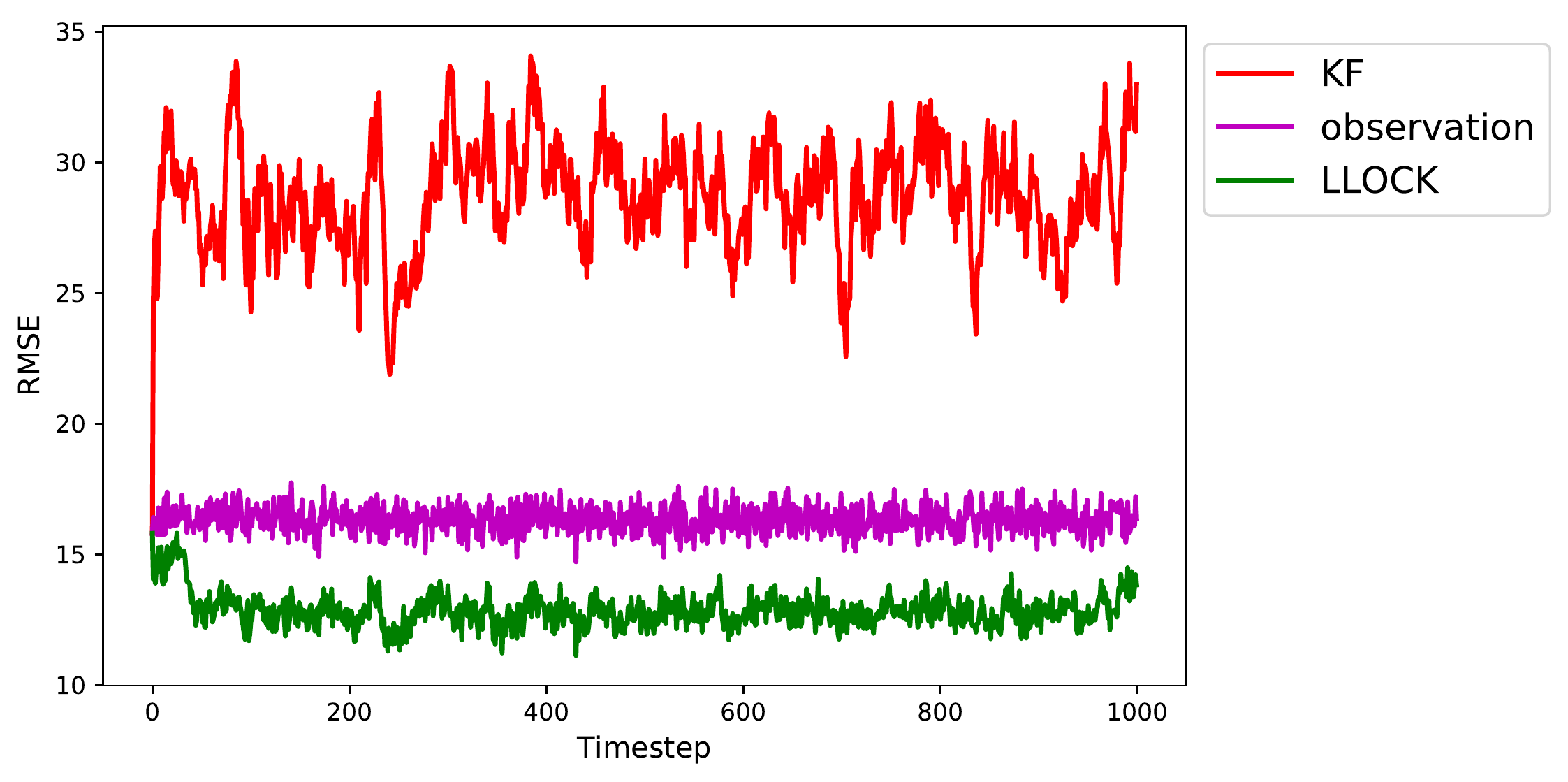}{Time transition of the RMSE of the observations and estimated results used by KF and LLOCK.}{fig:lsf_rmse}{12}
\figimage{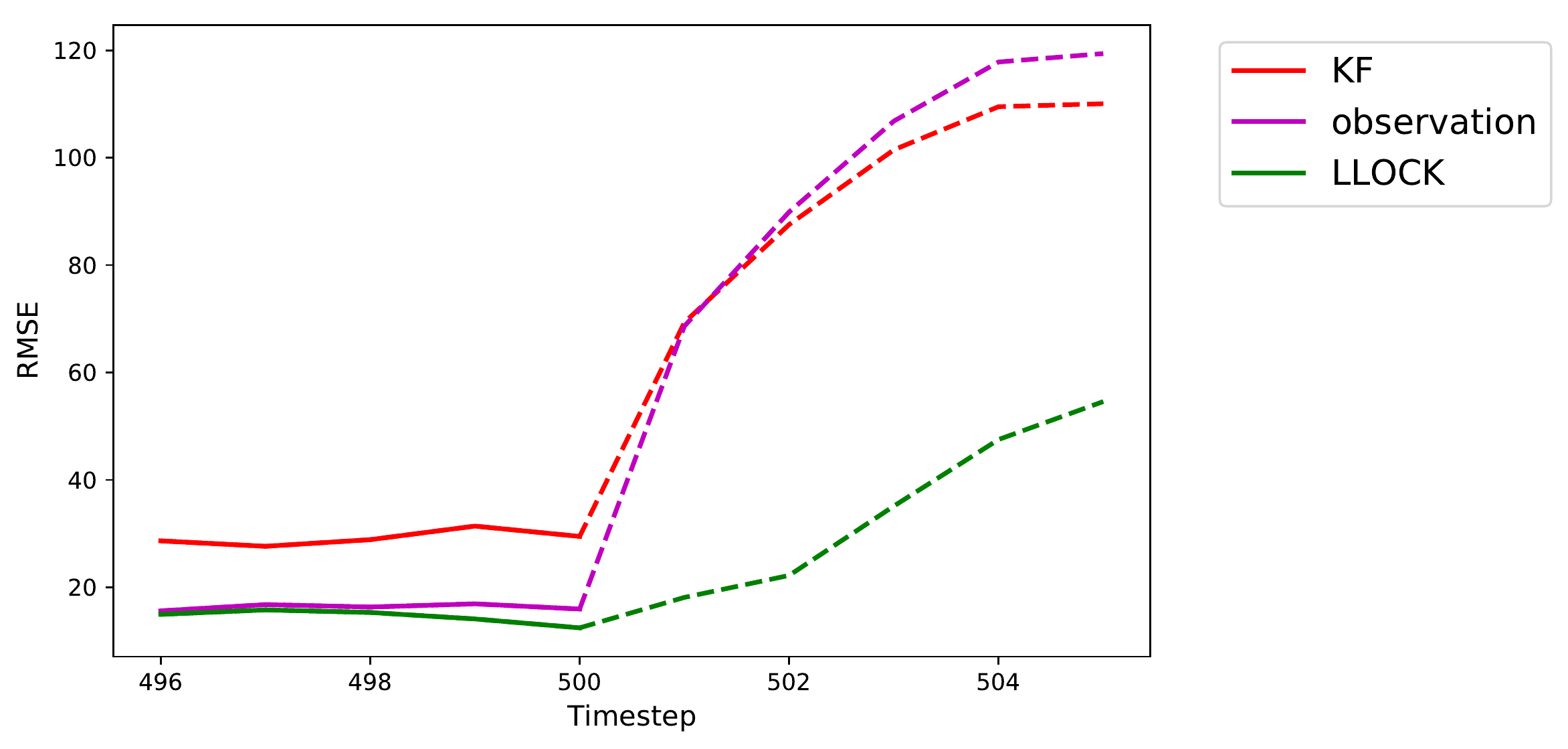}{Time transition of the RMSE between the true states and the observations and the predicted results used by KF and LLOCK when we obtain data until $t=500$ and perform prediction thereafter.}{fig:lsf_stp}{12}

\section{CONCLUSION}
\label{sec:conclusion}
In the present paper, we propose three real-time methods to estimate the states and state transition matrices in a linear Gaussian state space model. The first proposed method, namely, linear operator construction with the Kalman filter (LOCK), can approach the true transition matrices and the true states via application of the damped oscillation model. The advance methods, namely, SLOCK and LLOCK, achieve better performance in terms of noise reduction and short-term prediction through the three synthetic data: object moving, global flow, and local stationary flow. These methods are also superior to the EM algorithm in terms of computational and memory cost. In fact, the calculation time and memory usage of these methods are much less than those of the existing method. Therefore, these methods have the potential to estimate the transition of data, such as weather forecasting and object tracking, in real time.

Nevertheless, the proposed methods have three main drawbacks: dependence on a linear Gaussian framework, tuning of hyper-parameters, and tight assumption of SLOCK. First, as the proposed methods use a linear Gaussian formulation, we cannot directly apply these methods to nonlinear or non-Gaussian data. Then, the proposed methods need to provide interpretability for such data. Second, the proposed methods include the hyper-parameters $\tau$, $\eta$, and $c$. An automatic tuning method for these parameters is necessary for application to real data. Finally, although SLOCK has better performance for spatially uniform data, this assumption is rather limiting for real data. In order to overcome this issue, we are eager to develop a combined method of LLOCK and SLOCK because these assumptions and precisions are in trade-off relationships.

\section*{ACKNOWLEDGMENT}
This work was supported in part by JST, PRESTO, JPMJPR1774 and Meiji University Special Research Project.

\clearpage
\appendix
\section{DAMPED OSCILLATION MODEL}
\label{sec:dom}
Here, we present the residual results of the damped oscillation model. Figures \ref{fig:dom_lock_mat} and \ref{fig:dom_emkf_mat} show the estimated transition matrices for experiment 2 by LOCK and EMKF, respectively. From these figures, the estimated elements of LOCK approach the true elements, whereas the correspondence of the EMKF method does not.

In addition, Figures \ref{fig:cdom_lock_mat001} and \ref{fig:cdom_emkf_mat001} correspond to experiment 5. Figure \ref{fig:cdom_lock_mat001} indicates that the estimated matrices by LOCK can match the true matrix, excluding the last few updates, the noise ratio of which are higher than in other matrices. Figure \ref{fig:cdom_emkf_mat001} indicates that the 100 simulated results are similar in terms of the time transition of the matrices.

\figimage{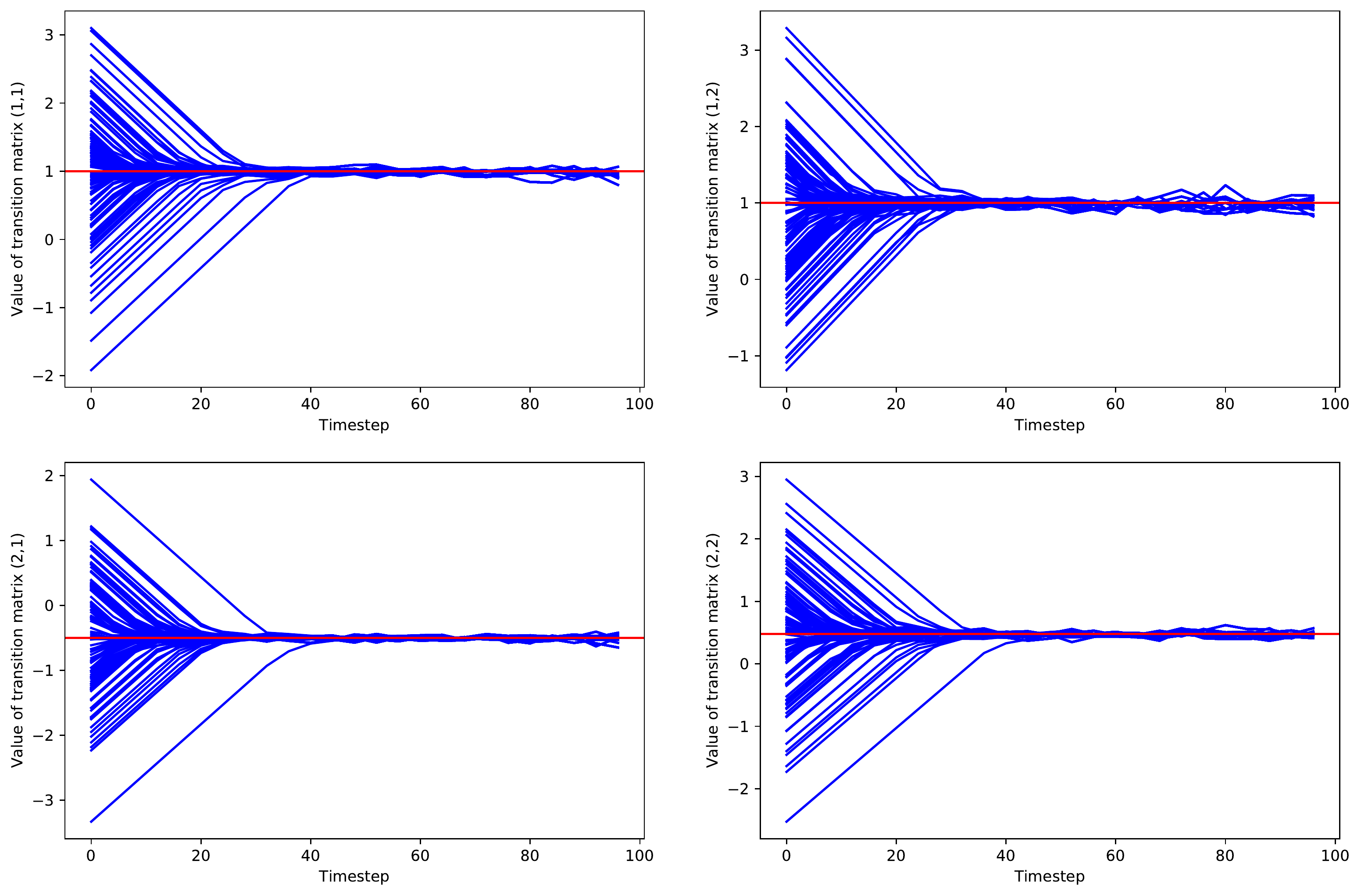}{Time transition of true transition matrices ({\it red}) and estimated transition matrices by LOCK ({\it blue}) in experiment 2. The four figures correspond to four elements of the matrices. For instance, the upper-left panel corresponds to $F_{11}$.}{fig:dom_lock_mat}{11}
\figimage{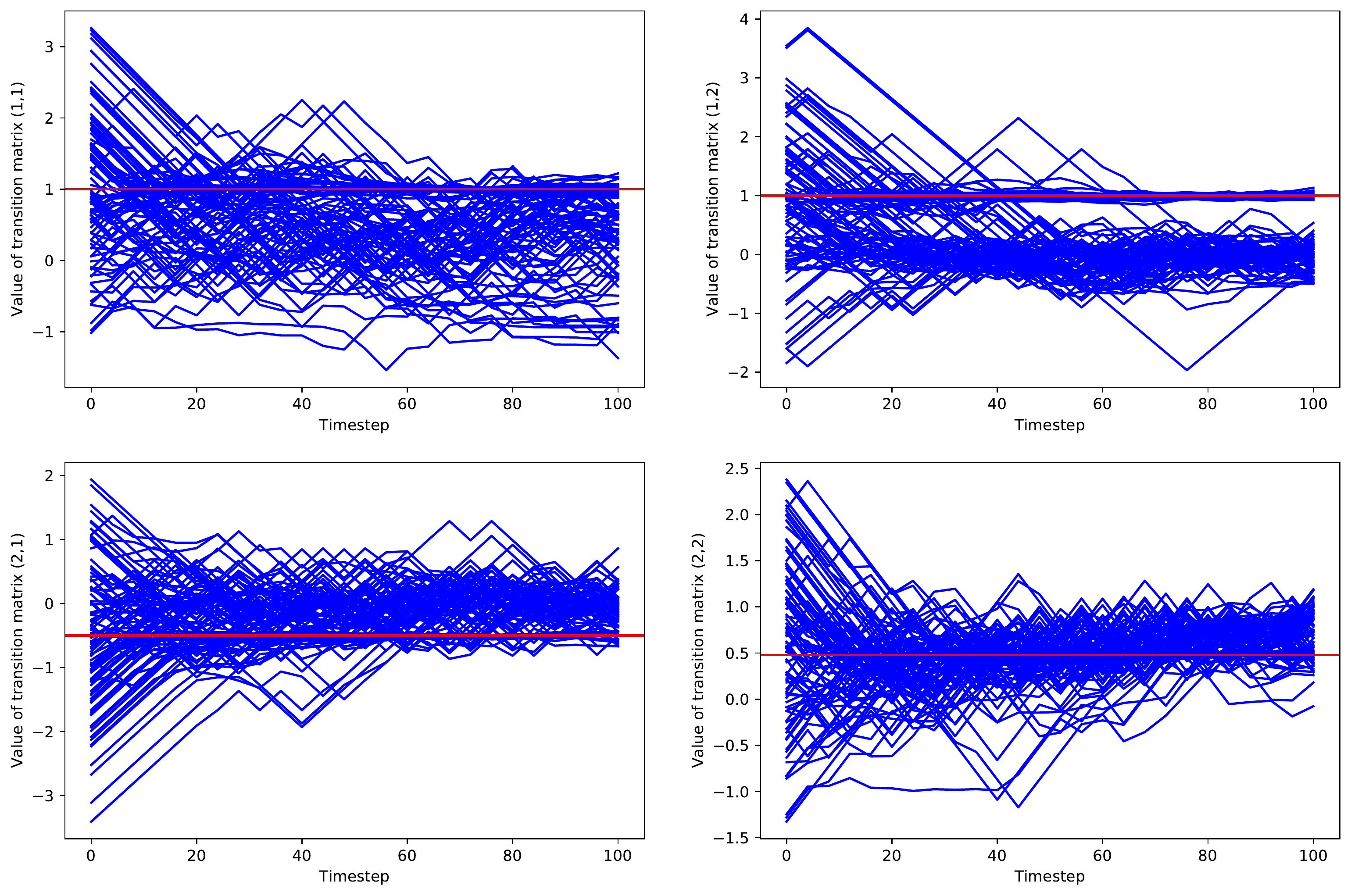}{Time transition of the true transition matrices ({\it red}) and the estimated transition matrices by EMKF ({\it blue}) in experiment 2. The four panels correspond to the four elements of the matrices. For instance, the upper-left panel corresponds to $F_{11}$.}{fig:dom_emkf_mat}{11}
\figimage{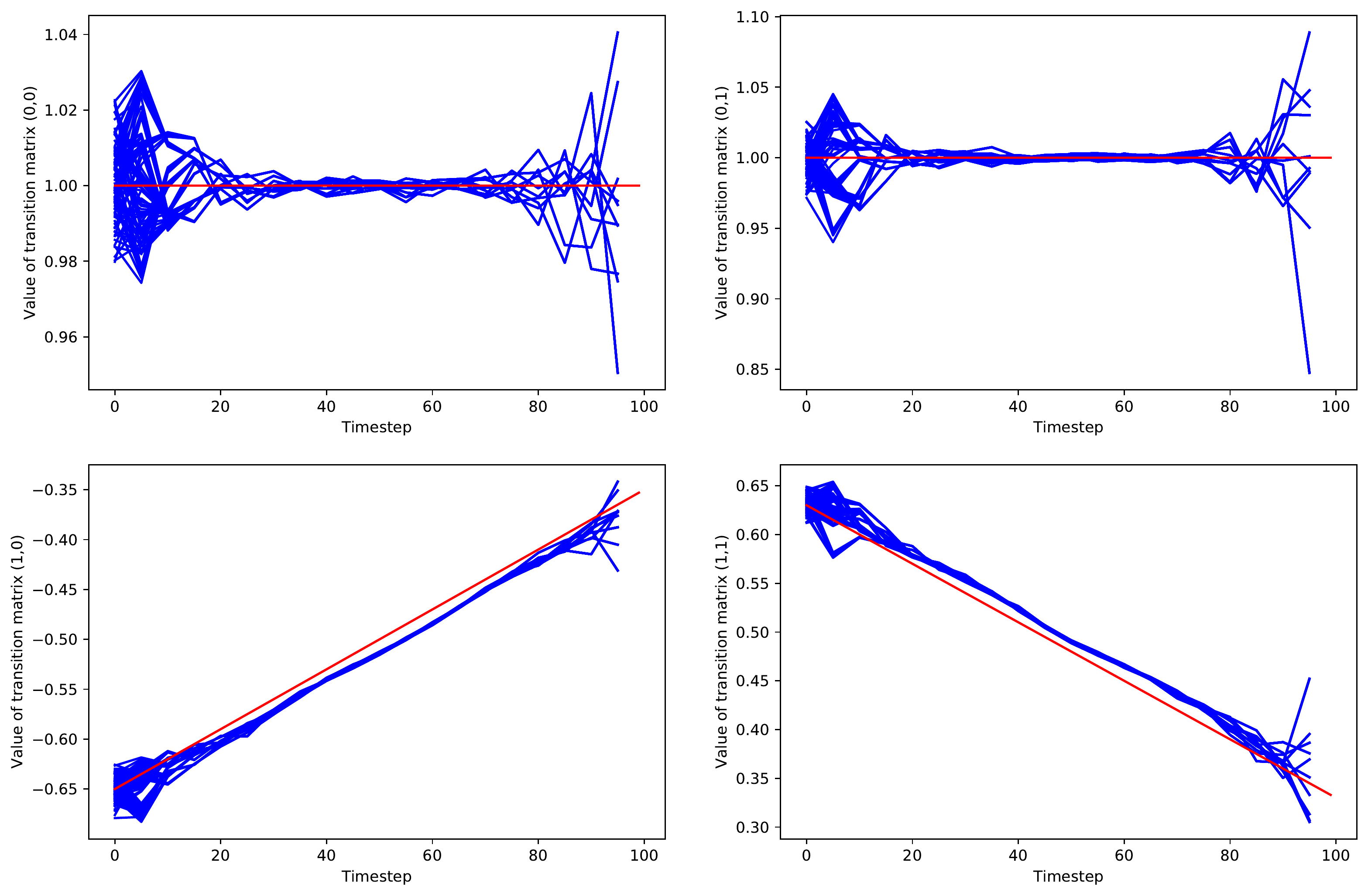}{Time transition of the true transition matrices ({\it red}) and the estimated transition matrices by LOCK ({\it blue}) in experiment 5. The four panels correspond to the four elements of the matrices. For instance, the upper-left panel corresponds to $F_{11}$.}{fig:cdom_lock_mat001}{11}
\figimage{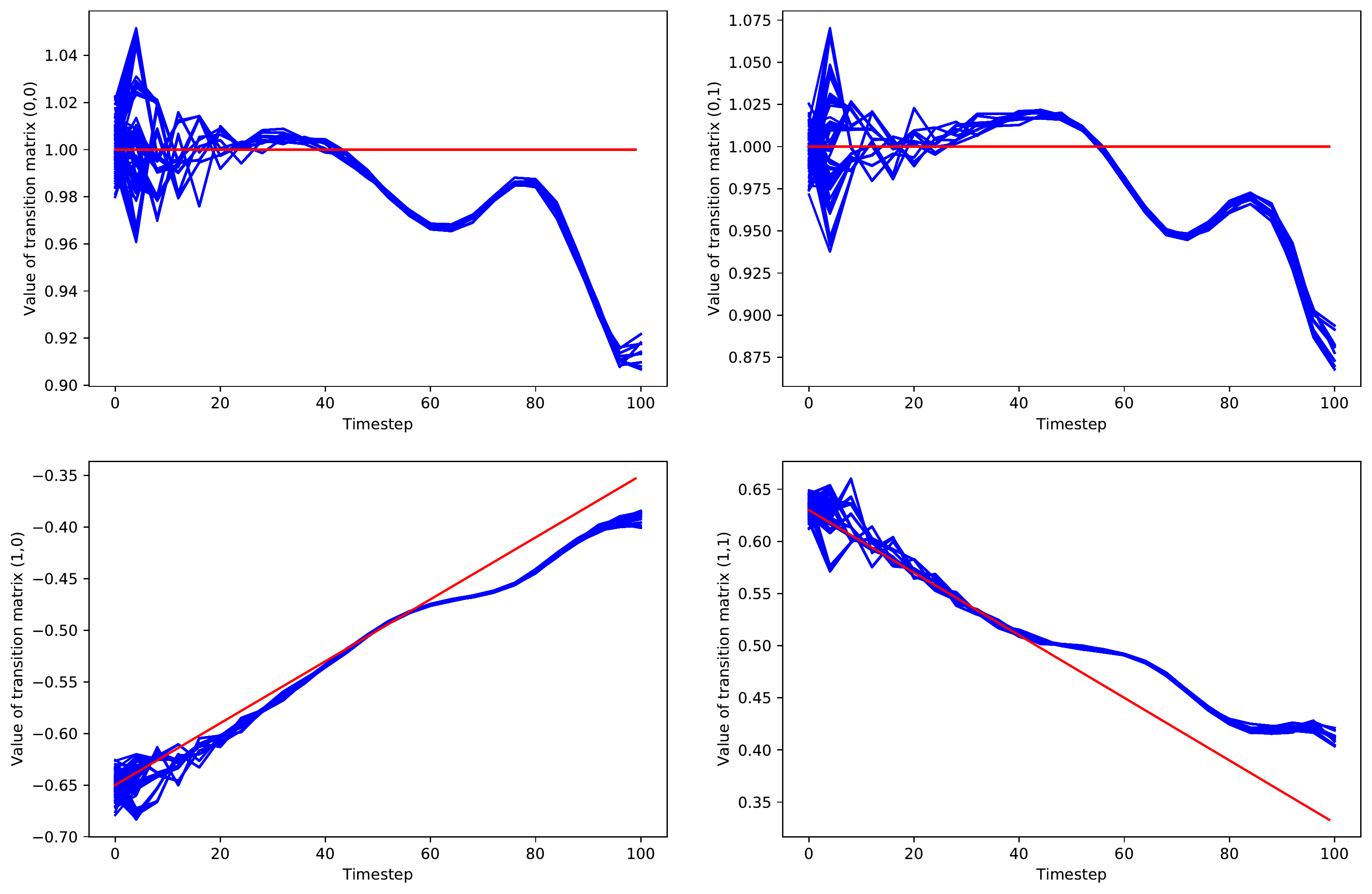}{Time transition of the true transition matrices ({\it red}) and the estimated transition matrices by EMKF ({\it blue}) in experiment 5. The four panels correspond to the four elements of the matrices. For instance, the upper-left panel corresponds to $F_{11}$.}{fig:cdom_emkf_mat001}{11}

\section{OBJECT MOVING DATA}
\subsection{How to Generate Data}
\label{ssec:mt_data}
First, we constructed a $25\times25$ array, the elements of which are 20. Then, we created random $N=15$ core points, the $x$ and $y$ coordinates of which exist from 7 to 16. Second, we selected two random core points from the points and generated a link, which links the two points, the width of which is $w=2$. Third, we assigned the coordinates in the link linear $z$ values from $(V_{max}-V_{min})\times m/M$, added a Gaussian noise, mean and standard deviation of which are both 10, where $V_{min}=100$, $V_{max}=150$, $m$, and $M$ are the minimum $z$ value, the maximum $z$ value, the iteration number, and the number of iterations, respectively. We iterated the second and third processes for $M=10$ times and obtained the base true image as shown in Figure \ref{fig:mt_init}. 

Fourth, we set the direction list and direction change list as shown in Figure \ref{fig:mt_data}. In each direction, we constructed the translation matrix regarding the direction and then produced the image vector and the matrix. As a result, we obtained the true data. Finally, we added a Gaussian noise having a mean and standard deviation of 0 and 20, respectively, to the true data. We applied the absolute operator to the noise because the image data are greater than 0. Figure \ref{fig:mt_init} shows images of the true data and the pseudo-observation data. This generating process is also shown on our GitHub page.

\figimagelf{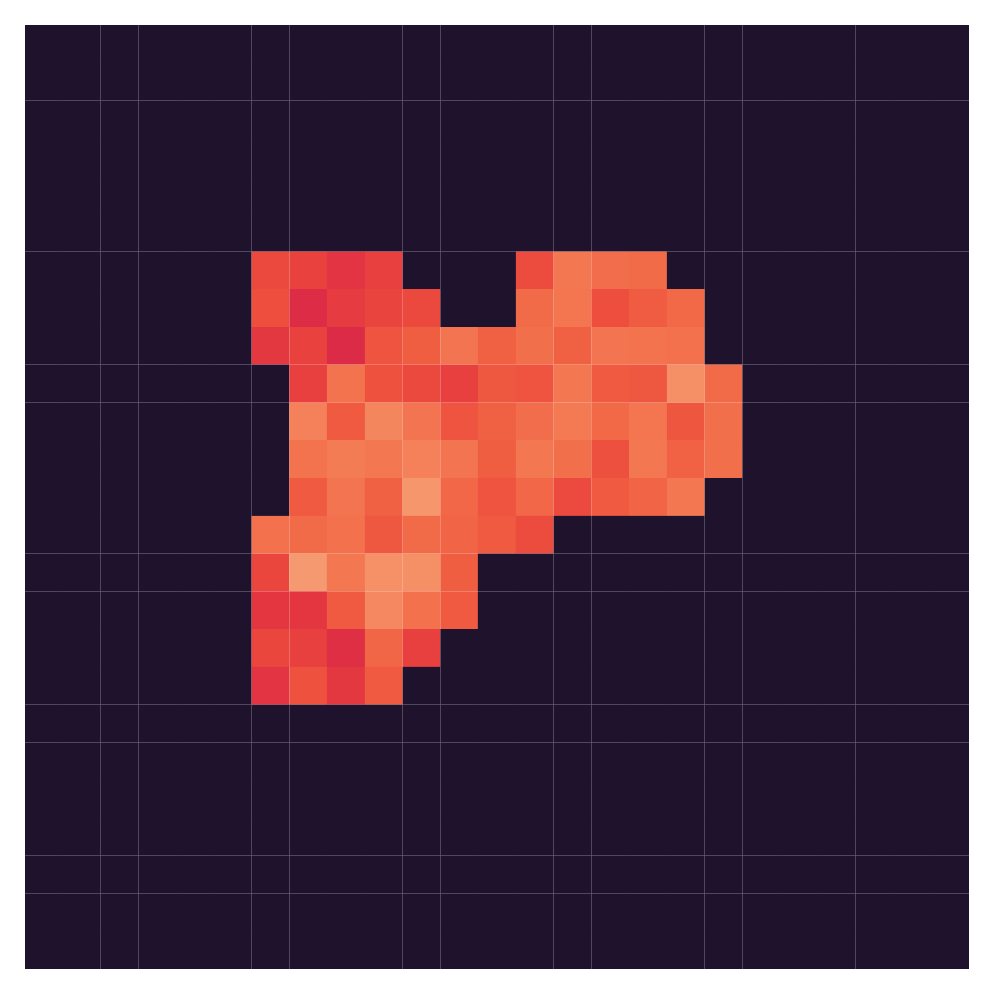}{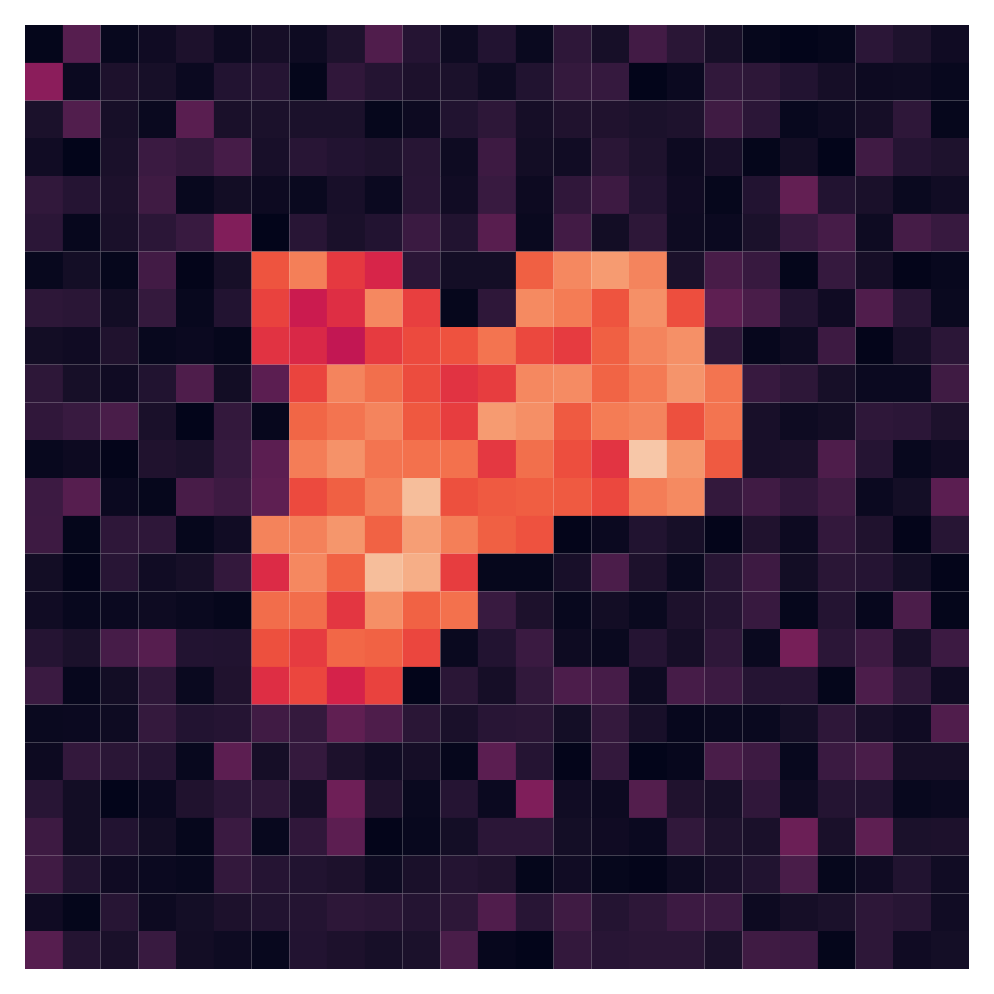}{(Left) True image of the object moving data. (Other) Observation data at the initial time.}{fig:mt_init}{5}{5}

\subsection{Residual Results}
\label{ssec:mt_res}
Figure \ref{fig:mt_srmse} shows the substantial RMSE (SRMSE) between the estimated transition matrices and the true transition matrices formulated by
\begin{equation}
SRMSE(F,\hat F)=\sqrt{\frac{1}{n_z}\sum_{(i,j)\in\{(i,j)|L_{ij}>0\lor P_{ij}>0\}}(F_{ij}-\hat F_{ij})^2}\\
\end{equation}
where $n_z$ represents the number of non-zero elements of the localization matrix, i.e., 
\begin{equation}
n_z=\sum_{i,j=1}^lL_{i,j}=\sum_{i,j=1}^l\delta(P_{i,j}>0).
\end{equation}
From this figure, the estimated matrices are close to the true matrix, especially the elements having true values of zero.

In addition, we conducted sensitivity analysis as shown in Figure \ref{fig:mt_sa}. This figure indicates that the hyper-parameters of SLOCK are robust for this data.

\figimage{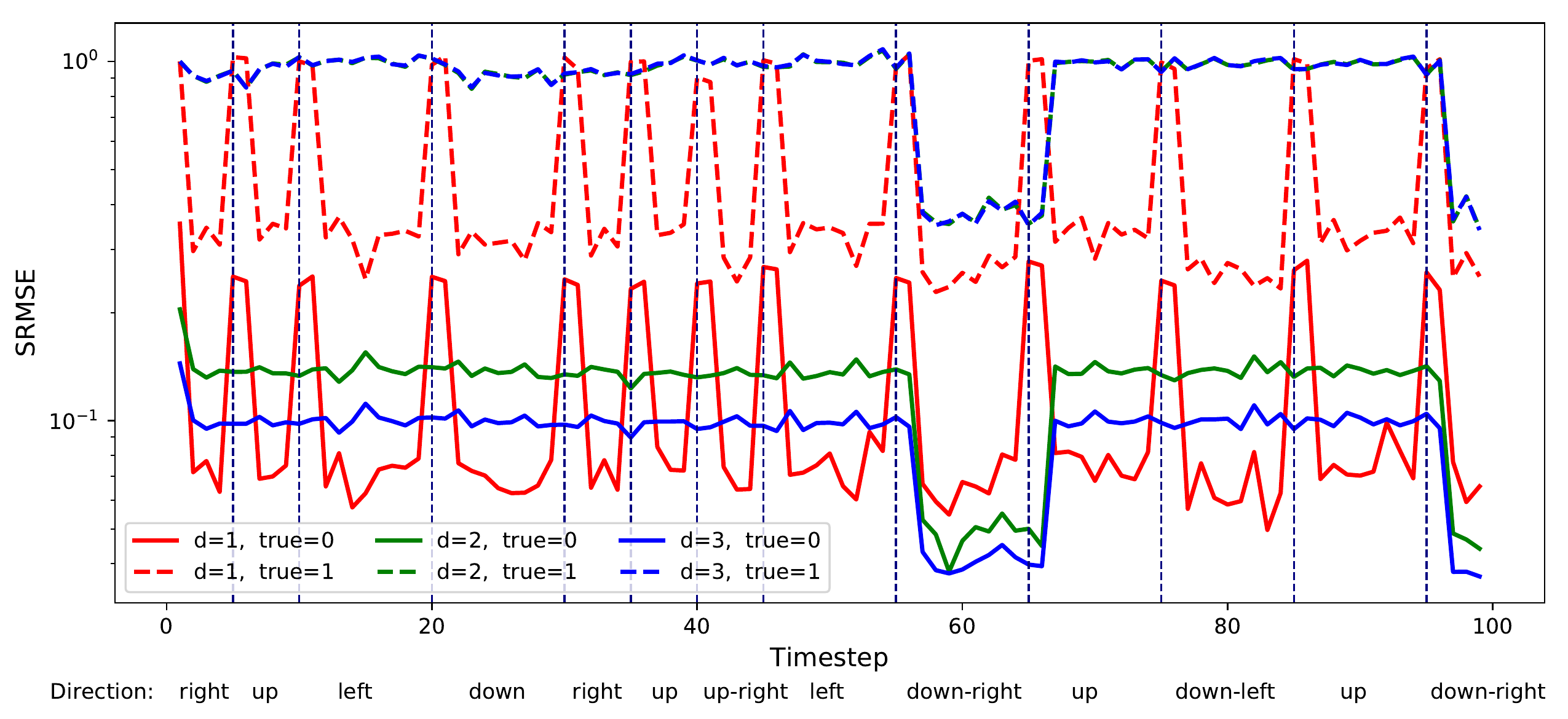}{Time transition of the SRMSE between the true transition matrices and the estimated correspondence. Here, ``true$=0$" indicates the calculation of the SRMSE for elements having true values that are zero.}{fig:mt_srmse}{12}
\figimagethree{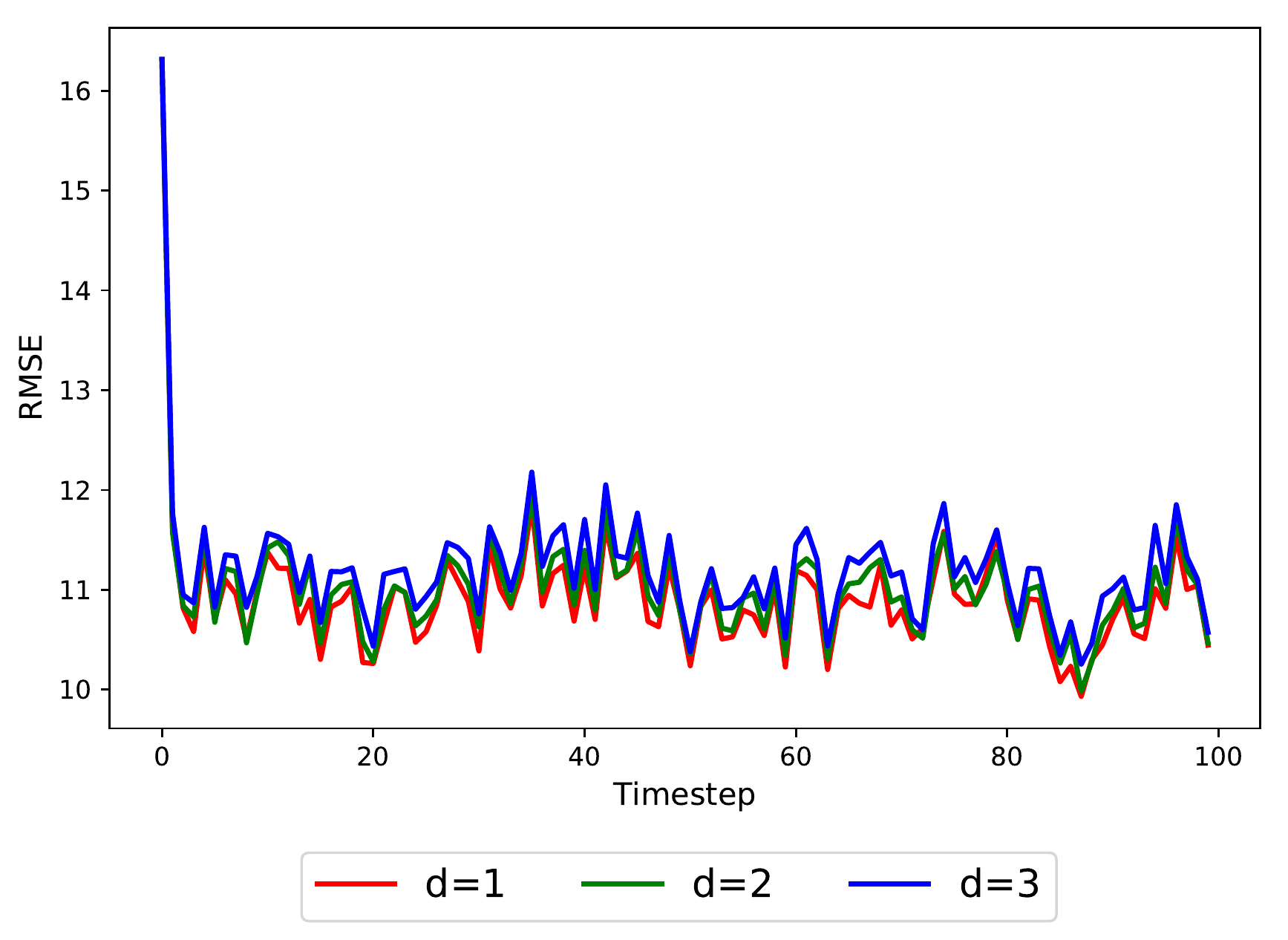}{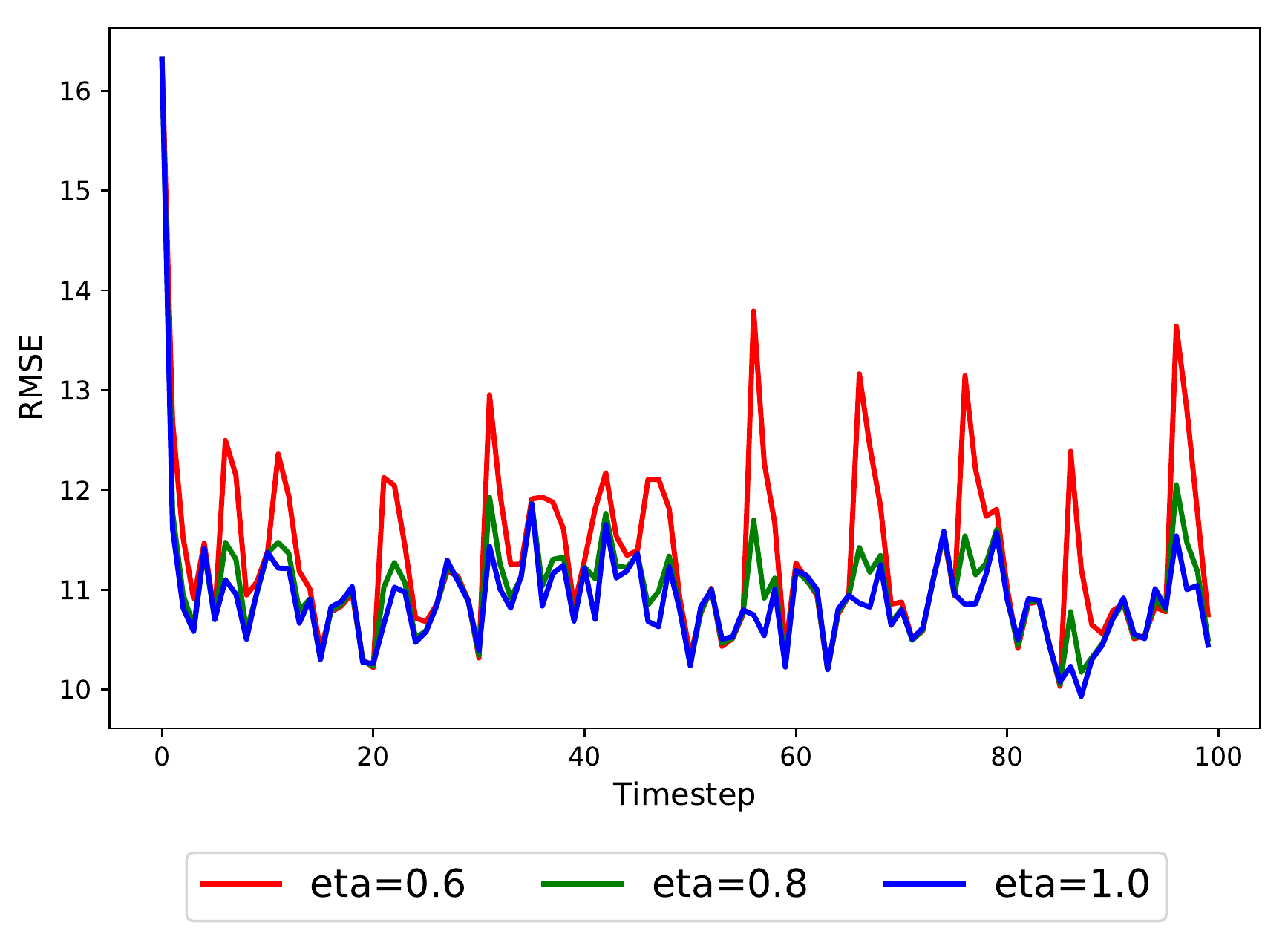}{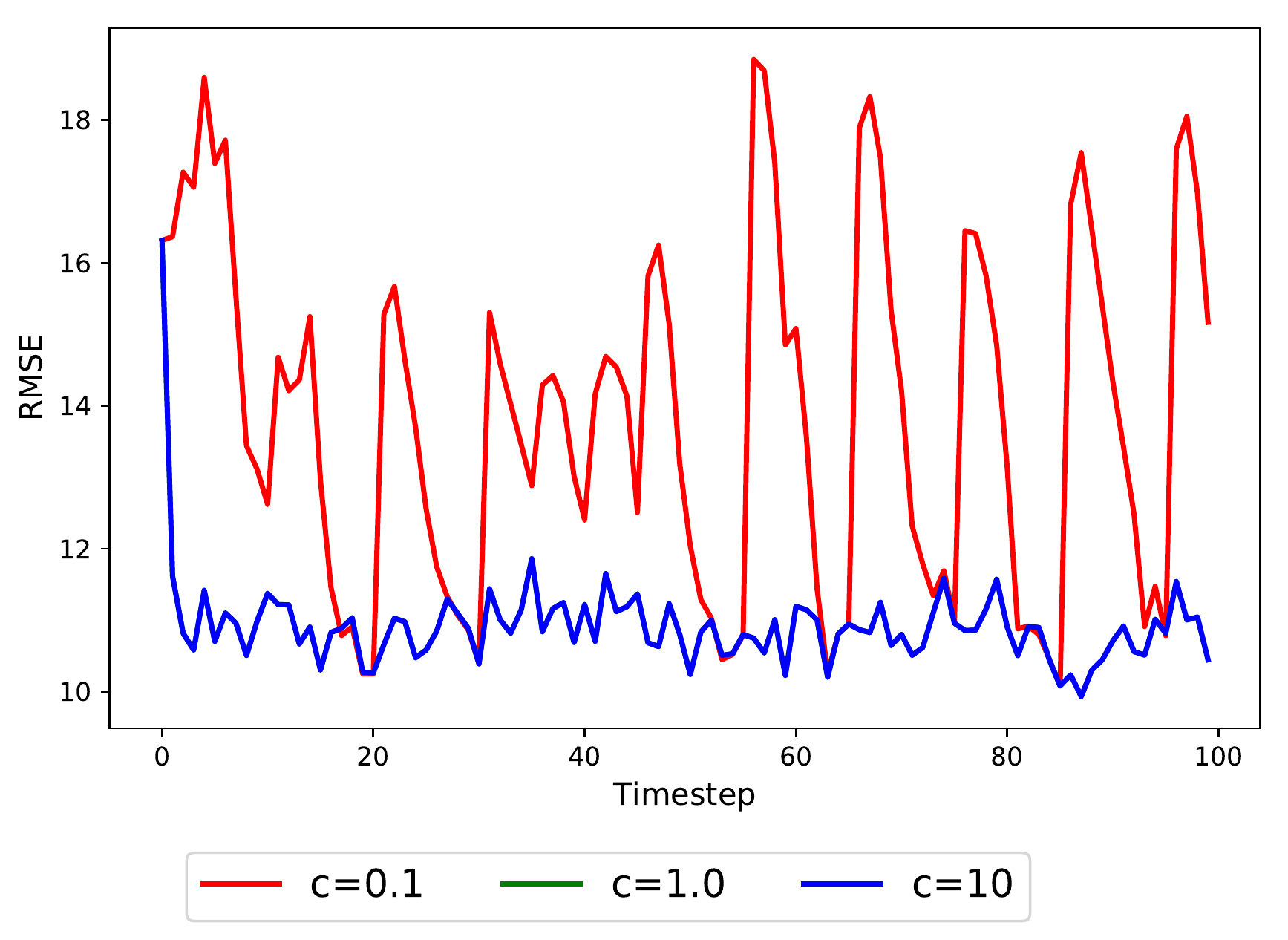}{Sensitivity analysis regarding adjacency distance $d$, learning rate $\eta$, and cutoff distance $c$. During analysis of the other parameters, we set $d=1$, $\eta=1.0$, and $c=1.0$.}{fig:mt_sa}

\section{GLOBAL FLOW}
\subsection{How to Generate Data}
\label{ssec:gf_data}
First, we made a $30\times30$ matrix, the elements of which are 20. Then, we created random objects, the sizes of which are randomly chosen from two to four and the values of which are i.i.d. $N(150,20^2)$. Second, we set the directions at each interval and  move the objects corresponding to the direction at each time interval. Finally, we added a zero-mean Gaussian noise, the standard deviation of which is 20. This generating process is also shown on our GitHub page.

\subsection{Residual Results}
\label{ssec:gf_res}
We calculated the SRMSE between the true transition matrices and the estimated matrices as shown in Figure \ref{fig:gf_srmse}. The results indicate the estimated matrices are close to the true matrices, excluding rapid change points.

In addition, Figure \ref{fig:gf_sa} represents the results of sensitivity analysis for $\tau$, $\eta$, and $c$. From this figure, LLOCK has robustness for the synthetic data.

\figimage{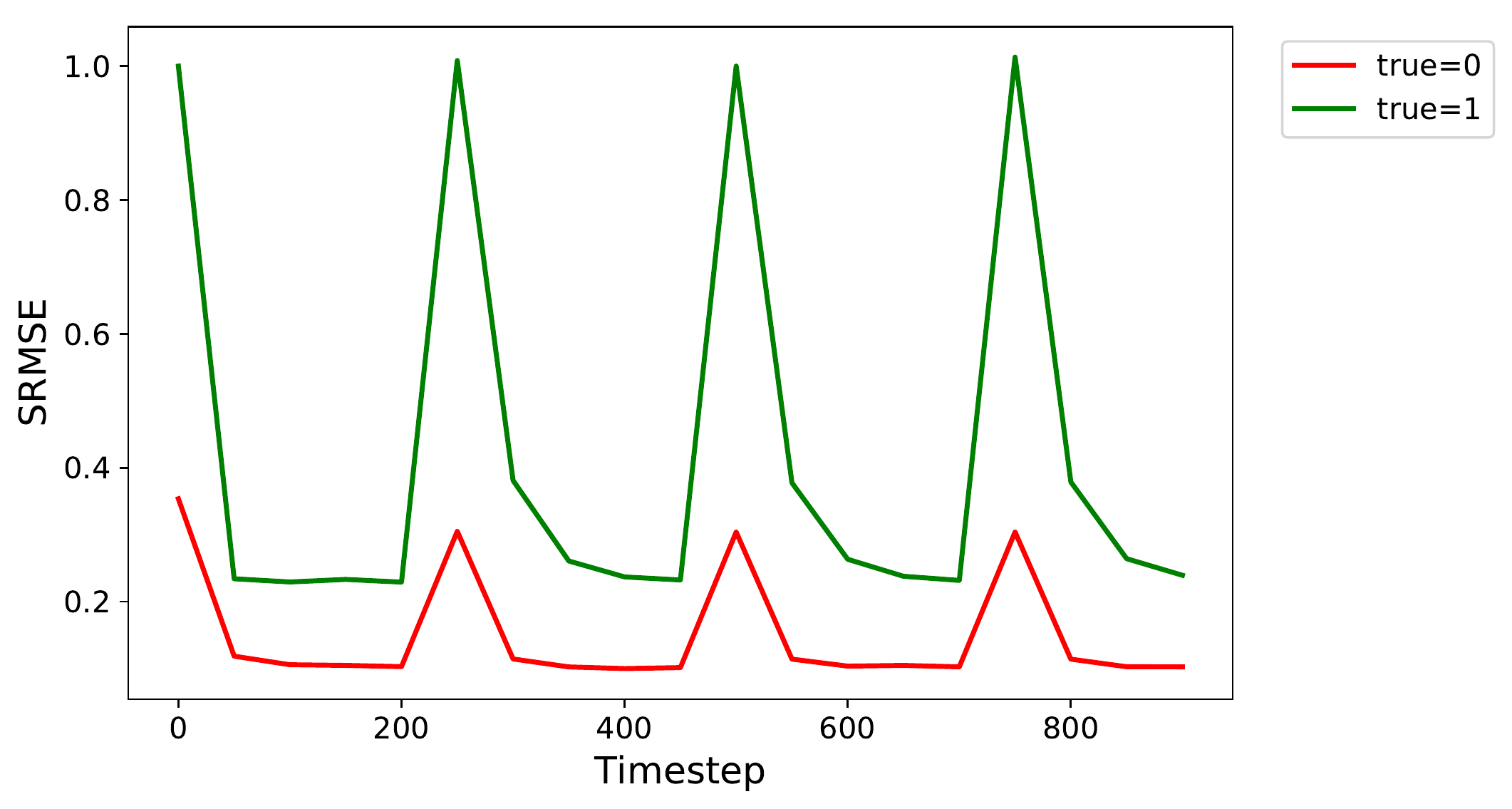}{Time transition of the SRMSE between the true transition matrices and the estimated correspondence. Here, ``true$=0$" indicates the calculation of the SRMSE for elements having true values of zero.}{fig:gf_srmse}{12}
\figimagethree{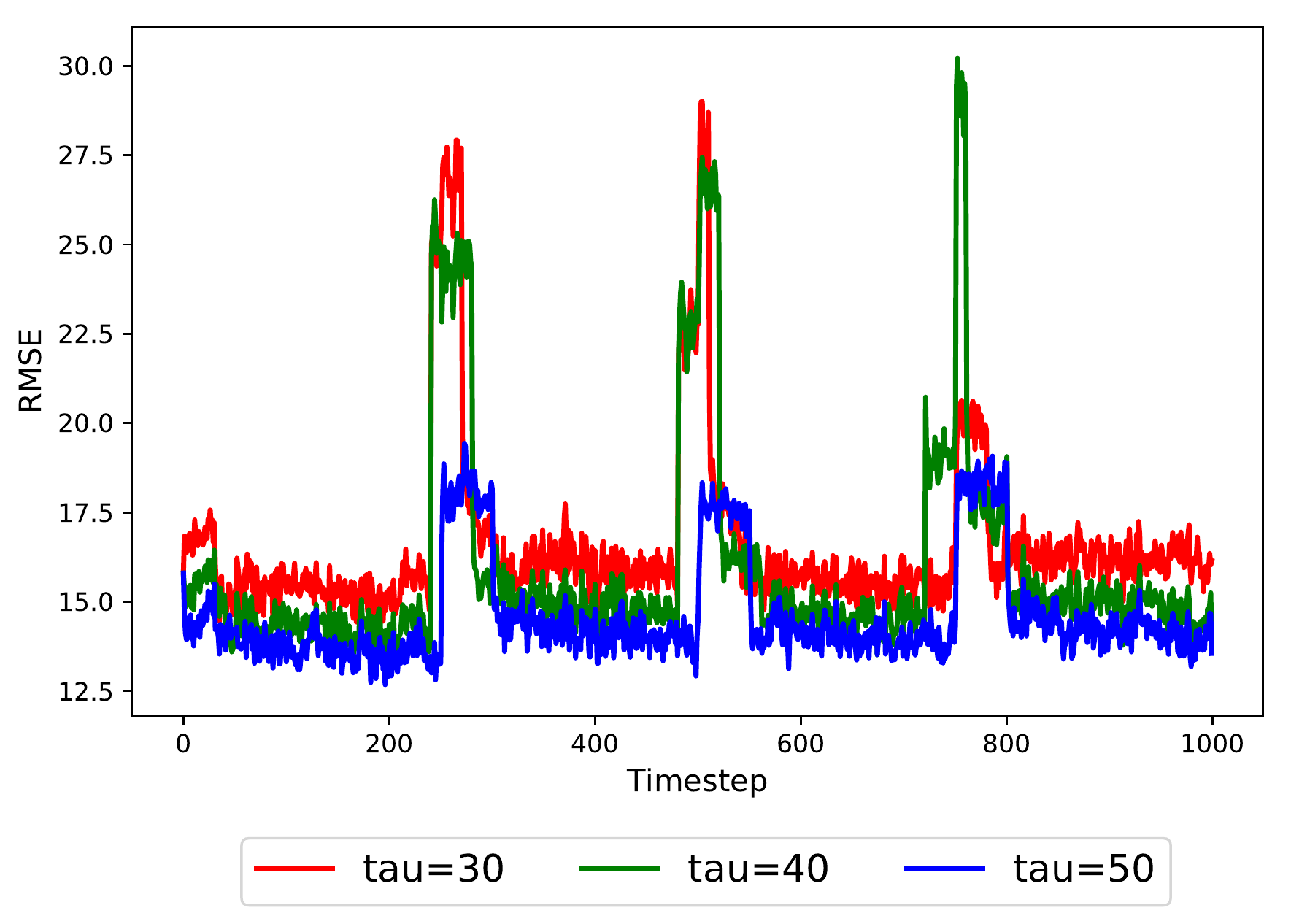}{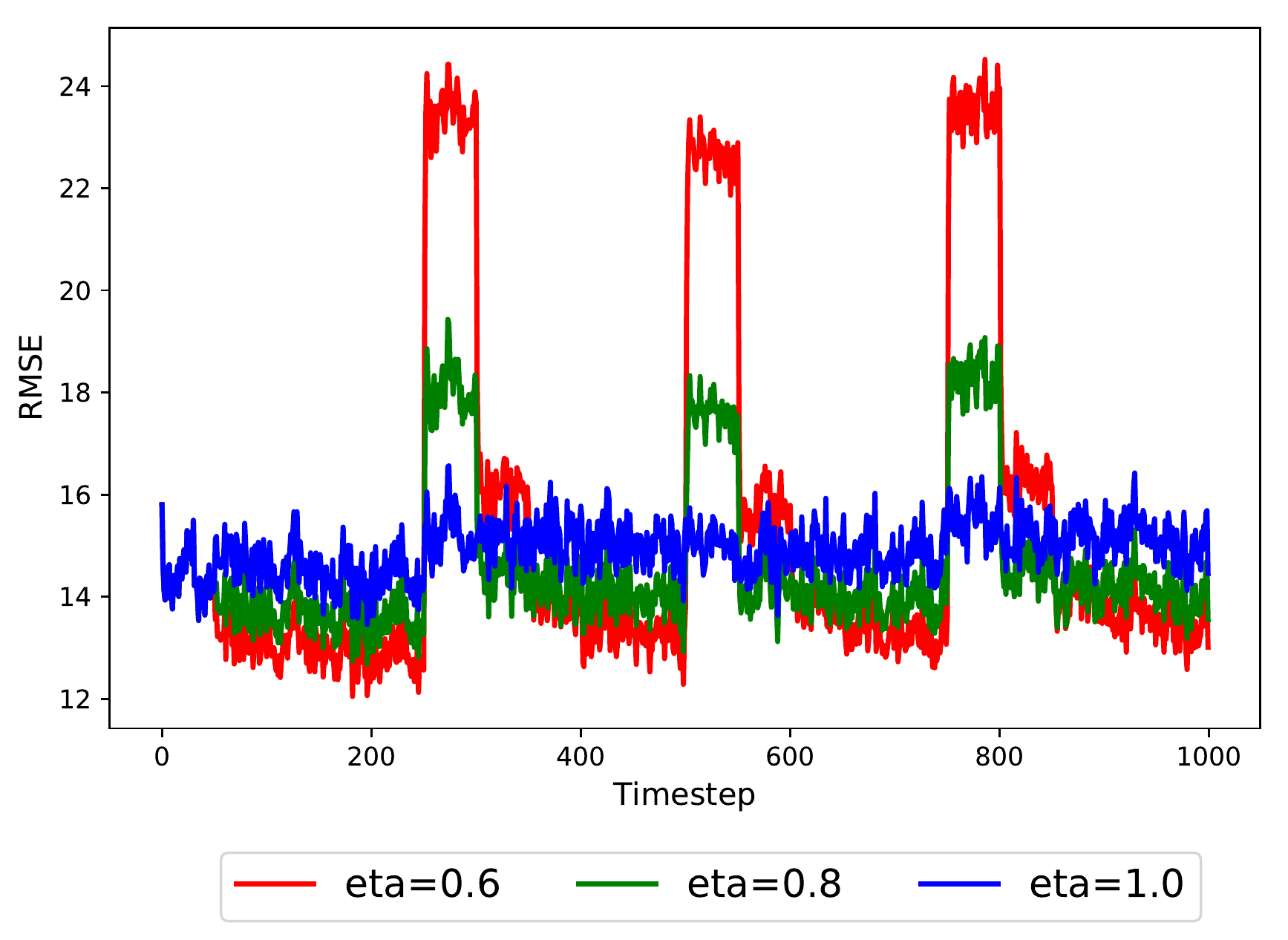}{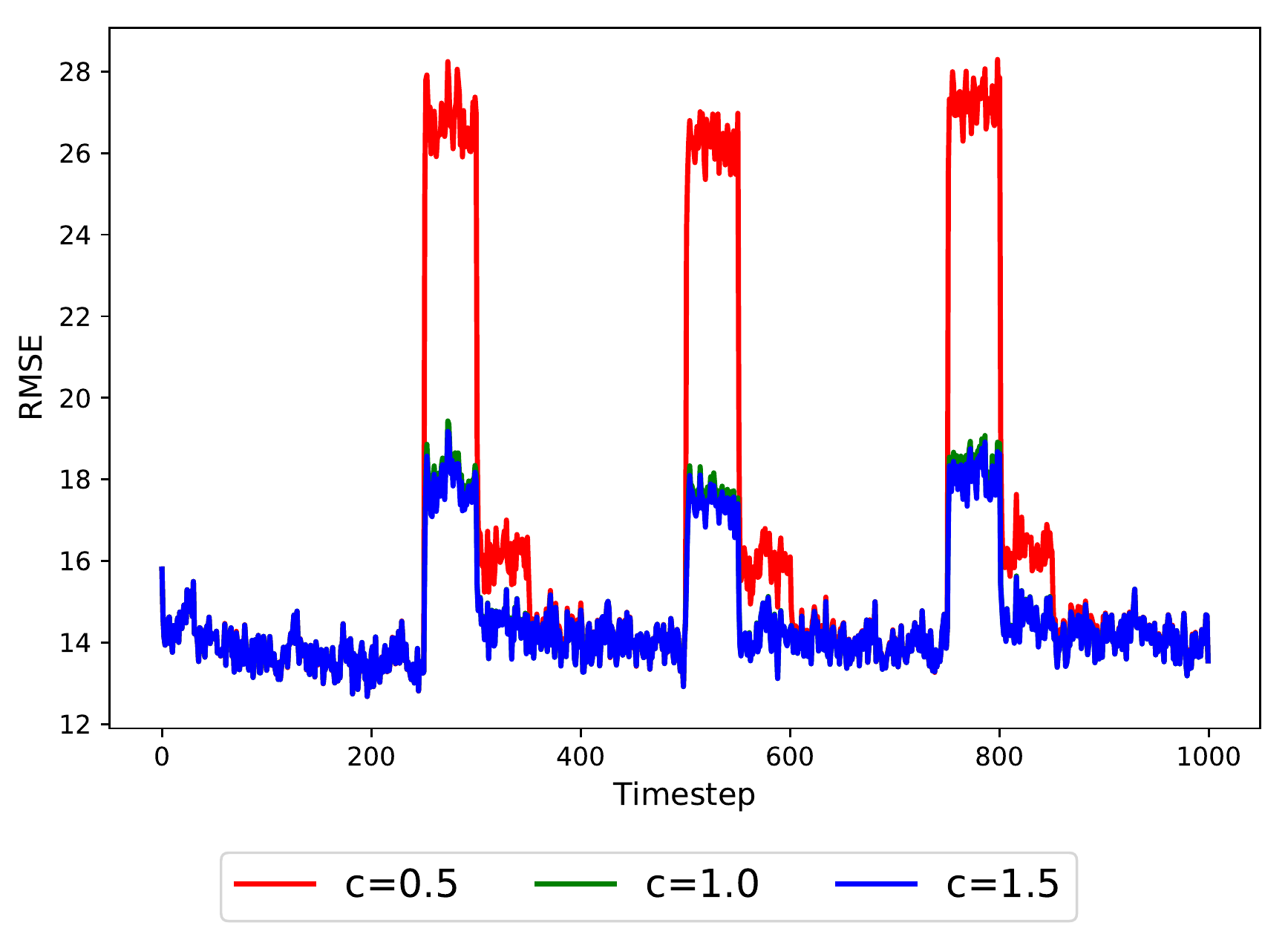}{Sensitivity analysis regarding update interval $\tau$, learning rate $\eta$, and cutoff distance $c$. During analysis of the other parameters, we set $\tau=50$, $\eta=0.8$, and $c=1.0$.}{fig:gf_sa}

\section{LOCAL STATIONARY FLOW}
\subsection{How to Generate Data}
\label{ssec:lsf_data}
First, we constructed a $(T+L-1)\times L$ matrix, the elements of which are 20, where $T=1000$ and $L=15$ indicate the number of time steps and the block length, respectively. Then, we created random objects, the sizes of which were randomly chosen from two to four and the values of which follow Gaussian distribution $N(150,20^2)$, that were substituted into the object values for the random elements of the matrix. Second, we set the directions of four square blocks, the size of which are $15\times15$, divided from a $30\times30$ field, as shown in Figure \ref{fig:lsf_data}. Then, we set the initial value from the first 15 columns of the source matrix. Third, we moved the objects following the block flow and set new source values from the $t+L$ column of the matrix. Finally, we added zero-mean Gaussian noise having a standard deviation of 20. This generating process is also shown on our GitHub page.

\subsection{Residual Results}
\label{ssec:lsf_res}
Figure \ref{fig:lsf_srmse} shows time transition of the SRMSE between the true transition matrices and estimated correspondence. This figure shows that the estimated results by LLOCK are close to the values of the true matrix.

In addition, we conducted sensitivity analysis for $\tau$, $\eta$, and $c$, as shown in Figure \ref{fig:lsf_sa}. This figure shows that LLOCK is a robust method for this local stationary flow data.

\figimage{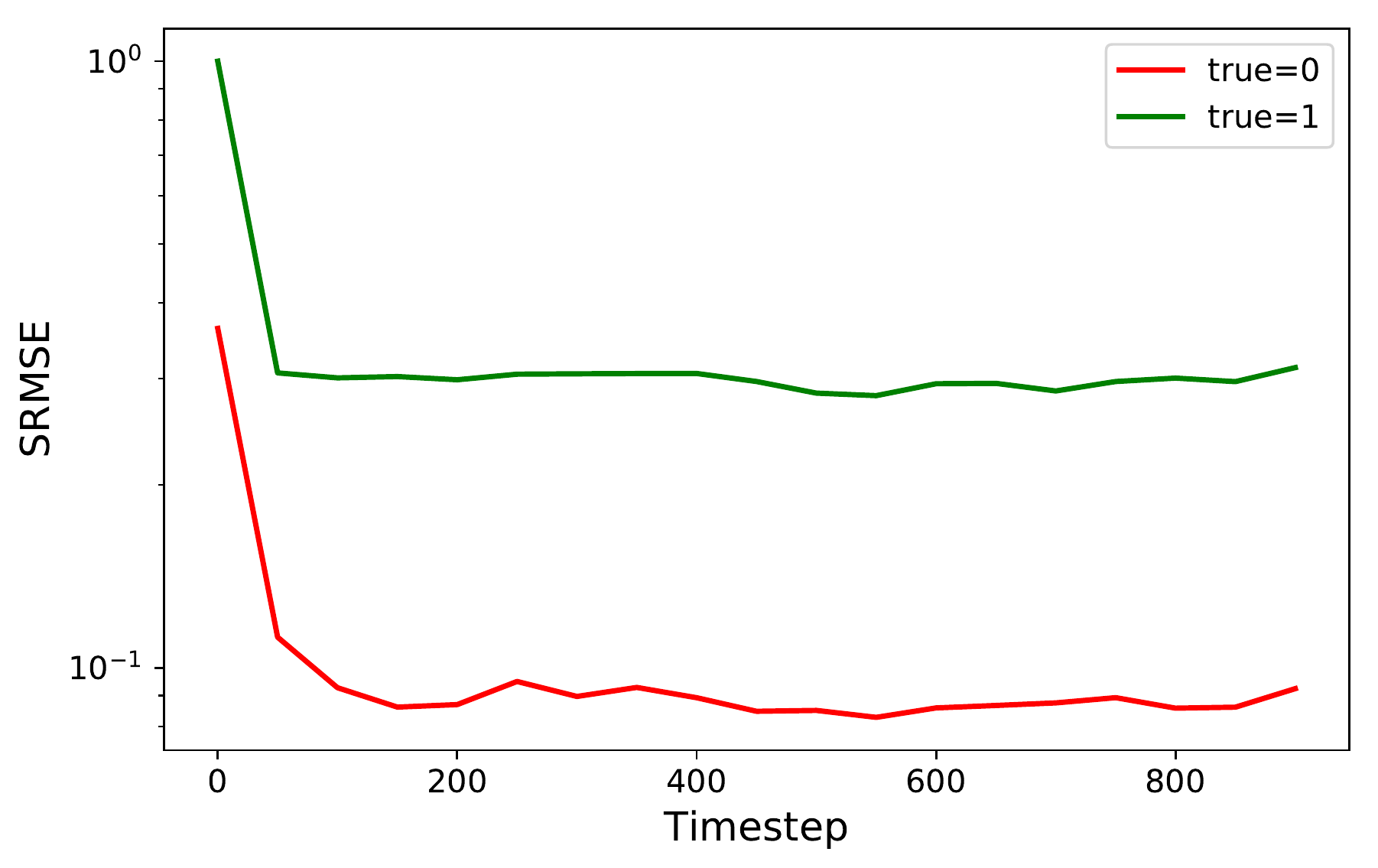}{Time transition of the SRMSE between the true transition matrices and the estimated correspondence. Here, ``true$=0$" indicates calculation of the SRMSE for elements having true values of zero.}{fig:lsf_srmse}{12}
\figimagethree{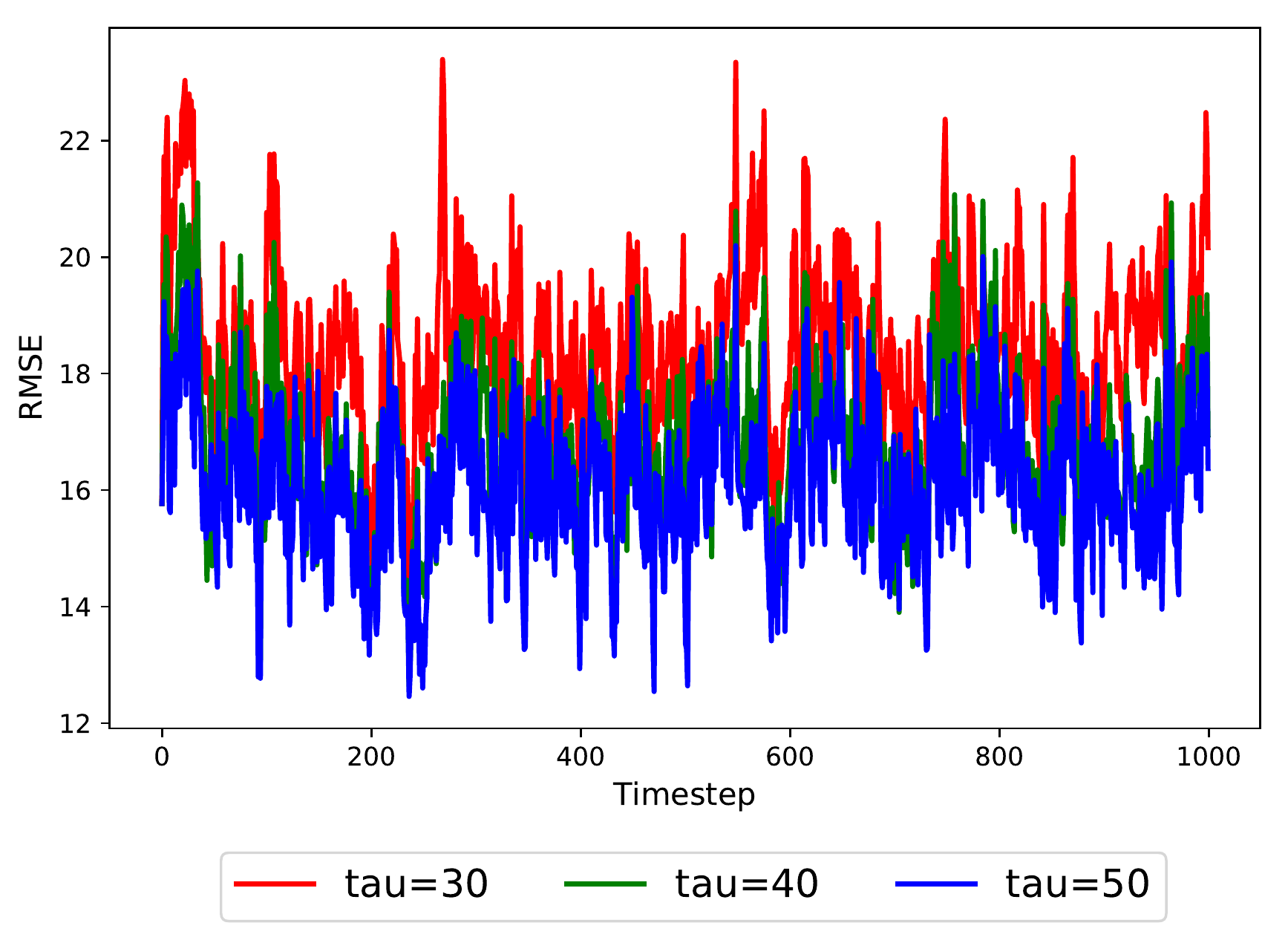}{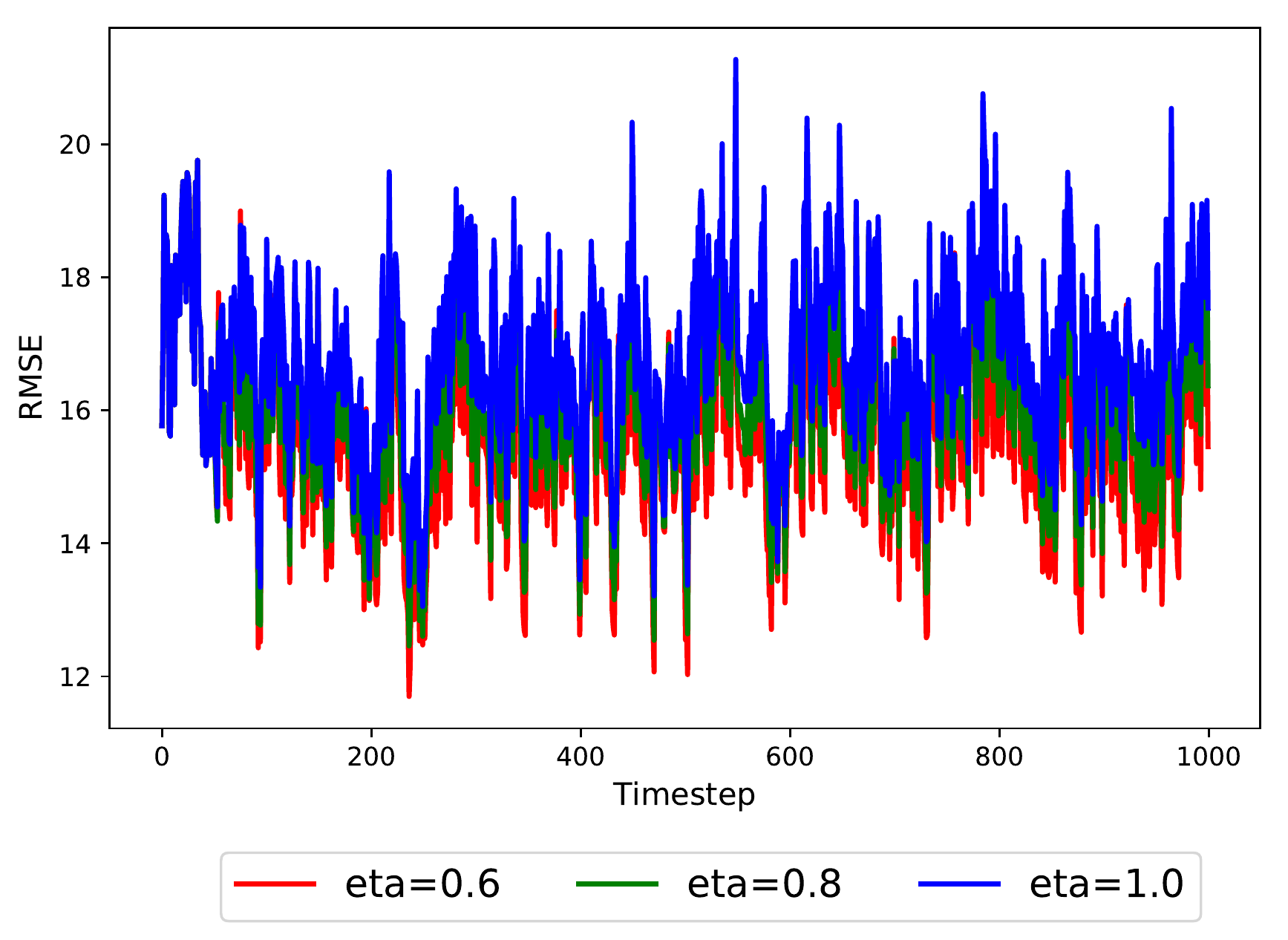}{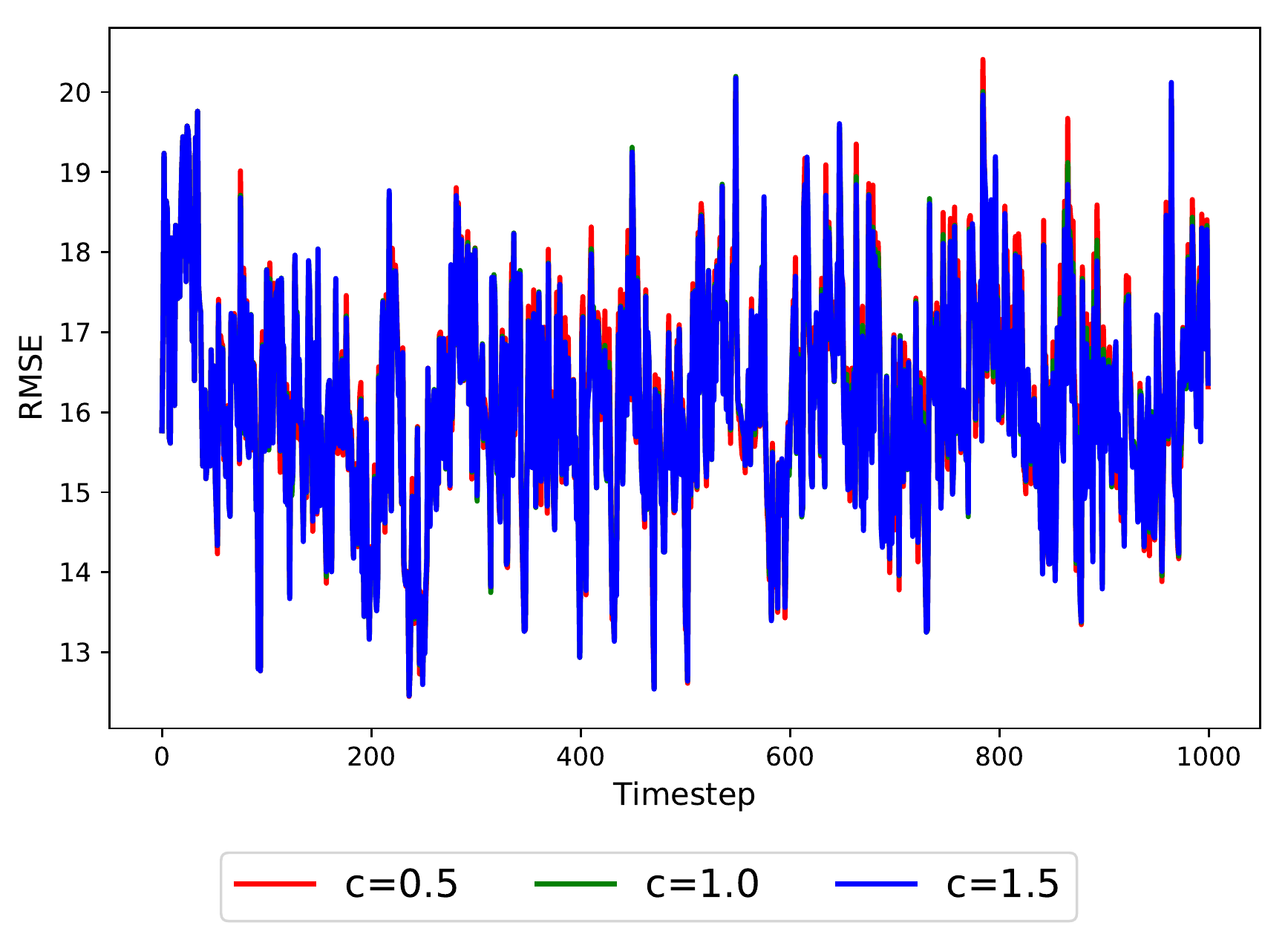}{Sensitivity analysis regarding update interval $\tau$, learning rate $\eta$, and cutoff distance $c$. During the analysis for other parameters, we set $\tau=50$, $\eta=0.8$, and $c=1.0$.}{fig:lsf_sa}

\section{COMPUTER ENVIRONMENT}
\label{sec:ce}
We used a GPU resource and Python. The following is the detailed environment of the present study.
\begin{itemize}
\item Calculation machine
\begin{itemize}
\item CPU: Intel Xeon E5-2670 2.6GHz (8core) x 2
\item Memory: 64 GB
\item HDD: SAS 300 GB $\times$ 2 (RAID 1)
\item OS: SuSE 12.0 Enterprise LINUX
\item GPU: NVIDIA Tesla K20, NVIDIA Quadro K5000
\item Other: CUDA8.0
\end{itemize}
\item Python
\begin{itemize}
\item Python 3.7.3
\item Anaconda 4.6.14
\item cupy-cuda80 5.4.0
\item numpy 1.16.2
\end{itemize}
\end{itemize}

\end{document}